\pdfoutput=1

\documentclass[11pt]{article}
\usepackage{longtable}
\usepackage[]{acl}
\usepackage{times}
\usepackage{latexsym}
\usepackage{booktabs}
\usepackage{amsmath, amssymb, graphicx, xcolor, multirow, multicol, comment, subcaption, url, float, etoolbox, adjustbox, pgf, soul, geometry, colortbl, booktabs}
\usepackage{amsmath}
\usepackage[ruled,vlined]{algorithm2e} 
\usepackage[T1]{fontenc}

\usepackage[utf8]{inputenc}

\usepackage{microtype}

\usepackage{inconsolata}

\usepackage{graphicx}
\usepackage{soul}
\usepackage{xcolor}
\newcommand{\fixme}[1]{{\color{red}\em\bf{[FIXME: #1]}}}

\usepackage{microtype}
\definecolor{lightred}{rgb}{1.0, 0.8, 0.8}
\definecolor{lightblue}{rgb}{0.8, 0.9, 1.0}
\definecolor{lightgreen}{rgb}{0.8, 1.0, 0.8}
\definecolor{lightyellow}{rgb}{1.0, 1.0, 0.8}
\definecolor{lightpurple}{rgb}{0.9, 0.8, 1.0}
\definecolor{lightorange}{rgb}{1.0, 0.9, 0.8}

\title{Multi-View Multi-Task Modeling with Speech Foundation Models for Speech Forensic Tasks}

\author{
  Orchid Chetia Phukan$^{1*}$, 
  Devyani Koshal$^{1}$\thanks{\footnotesize{ Authors contributed equally as first authors}} ,
  \textbf{Swarup Ranjan Behera}$^{2}$\\  
  \textbf{Arun Balaji Buduru}$^{1}$, 
   \textbf{Rajesh Sharma}$^{1,3}$\\
  \textsuperscript{1}IIIT-Delhi, India, 
  \textsuperscript{2}Reliance Jio AICoE, India, 
  \textsuperscript{3}University of Tartu, Estonia\\
  \texttt{\textbf{Correspondence:} \textcolor{blue}{(orchidp,devyani20055)@iiitd.ac.in}}
}

\begin{document}
\maketitle
\begin{abstract}

Speech forensic tasks (SFTs), such as automatic speaker recognition (ASR), speech emotion recognition (SER), gender recognition (GR), and age estimation (AE), find use in different security and biometric applications. 
Previous works have applied various techniques, with recent studies focusing on applying speech foundation models (SFMs) for improved performance. However, most prior efforts have centered on building individual models for each task separately, despite the inherent similarities among these tasks. This isolated approach results in higher computational resource requirements, increased costs, time consumption, and maintenance challenges. 
In this study, we address these challenges by employing a multi-task learning strategy. Firstly, we explore the various state-of-the-art (SOTA) SFMs by extracting their representations for learning these SFTs and investigating their effectiveness at each task specifically. Secondly, we analyze the performance of the extracted representations on the SFTs in a multi-task learning framework. We observe a decline in performance when SFTs are modeled together compared to individual task-specific models, and as a remedy, we propose multi-view learning (MVL). Views are representations from different SFMs transformed into distinct abstract spaces by characteristics unique to each SFM. By leveraging MVL, we integrate these diverse representations to capture complementary information across tasks, enhancing the shared learning process. We introduce a new framework called \textbf{TANGO} (\textbf{T}ask \textbf{A}lignment with I\textbf{N}ter-view \textbf{G}ated \textbf{O}ptimal Transport) to implement this approach. With \textbf{TANGO}, we achieve the topmost performance in comparison to individual SFM representations as well as baseline fusion techniques across benchmark datasets such as CREMA-D, emo-DB, and BAVED. 

\end{abstract}

\section{Introduction}
Forensic speech science is an indispensable discipline in criminal investigations and encompasses a wide range of applications, including security, user authentication, healthcare, customer service, and the entertainment industry, among others. Key tasks central to this field, such as Automatic Speaker Recognition (ASR), Speech Emotion Recognition (SER), Gender Recognition (GR), and Age Estimation (AE) - provide critical insights by analyzing paralinguistic features of speech, including pitch, intensity, tone, and other variations in speech. These tasks assist in identifying individuals, verifying claims based on vocal evidence, recognizing emotions for psychological assessments, enabling voice-based authentication, and facilitating demographic profiling for personalized services. 
As the prevalence of digital communication continues to surge, the imperative to extract intricate information from vocal data becomes increasingly pressing. This underscores the necessity for sophisticated models capable of concurrently addressing multiple forensic speech tasks. 

Early research on speech forensic tasks (SFTs) primarily relied on handcrafted features, such as MFCCs, and classical ML techniques like SVMs and k-NN \cite{abdulsatar2019age}. Although foundational, these methods often struggled with scalability and generalization across varied speech conditions. Neural Networks such as CNN and LSTM significantly enhanced the capacity to capture the complex acoustic and temporal dynamics inherent in speech signals \cite{zazo2018age}. This transition marked a substantial improvement, particularly in high-variability environments.

With the dawn of this decade, speech foundation models (SFMs) have transformed speech forensic analysis. SFMs such as  Wav2vec2, HuBERT, and WavLM generate robust, task-agnostic representations from raw speech, significantly improving performance across ASR, SER, GR, and AE \cite{yang2021superb, shor2022universal, lebourdais22_interspeech}. By leveraging large-scale pre-training, these SFMs exploit large volumes of data, capturing intricate paralinguistic features essential for forensic applications. Despite advancements in SFMs, their application in learning multiple SFTs simultaneously in a multi-task format remains underexplored. This multi-task learning strategy provides an efficient solution for saving computational, time, monetary, and maintenance challenges. Few prolific works have explored the use of SFMs in this direction \cite{zheng22b_interspeech, lee23g_interspeech}, however, there is a lack of comprehensive studies evaluating the effectiveness of various such SOTA SFMs across different SFTs, with no clear consensus on the best-performing models within a unified framework.



Additionally, SFMs vary significantly in design and training paradigms; some, like Wav2vec2 and WavLM, are self-supervised and learn representations from unlabelled data, while others, such as Whisper, are trained using labeled datasets. This diversity presents an opportunity for research to identify the most effective models for multi-task speech forensic analysis and optimize their integration within a unified framework. In response to these challenges posed, we conduct a detailed investigation for the first time, to the best of our knowledge, to determine the optimal SFM for concurrent training on ASR, SER, GR, and AE. Our analysis reveals that while the multi-task framework promises improved efficiency, it often leads to diminished performance when tasks are integrated due to task interference, underscoring shortcomings in current methodologies. This interference can be due to each individual SFT being dependent on different paralinguistic aspects of the input speech and the failure of individual SFM to effectively disentangle task-specific information.

To address this issue, we propose a novel approach known as multi-view learning (MVL), which combines representations from various SFMs\footnote{Here, each unique SFM representation is considered as view and used interchangeably with representation} and each SFM provides unique insights from their distinct abstract representations. This methodology facilitates the integration of diverse informational facets across tasks, thereby augmenting the comprehensive learning process and mitigating task interference. To our end, we introduce \textbf{TANGO} (\textbf{T}ask \textbf{A}lignment with I\textbf{N}ter-view \textbf{G}ated \textbf{O}ptimal Transport), a framework for effective MVL to synchronize these representations effectively. Our findings demonstrate that \textbf{TANGO} not only enhances performance compared to individual SFM outputs but also significantly surpasses baseline fusion methods on benchmark datasets, including CREMA-D, emo-DB, and BAVED.

\textbf{To summarize, the contributions of the work are as follows:} 
\begin{itemize}
    \item We present a comprehensive comparative analysis of SOTA SFMs for learning these SFTs individually. 
    \item We perform an investigative analysis of SOTA SFMs for learning the FSTs ASR, SER, GR, and AE in a multi-task learning format.
    \item Our findings highlight the performance trade-offs of jointly modeling these tasks, revealing significant challenges in current multi-task learning approaches due to the poor information disentanglement for each task through individual SFM representations.
    \item We introduce an MVL paradigm that integrates diverse representations from multiple SFMs, enhancing shared learning across tasks.
    \item We propose \textbf{TANGO} for aligning multi-view representations, outperforming baseline fusion techniques and individual SFMs.
\end{itemize}

The models and code developed for this study will be released after the review process.

\section{Related Works}

\label{section:second_section}

In this section, we will briefly discuss previous research on SFTs modeled using SFMs.

\noindent\textbf{Automatic Speaker Recognition}: \citet{shor2022universal} proposed universal paralinguistic conformer as a representation learning model for paralinguistic speech processing and showing SOTA performance for ASR. 
\citet{peplinski21_interspeech} leveraged TRILL to develop FRILL, which provides embeddings for paralinguistic applications in low-resource settings and demonstrates improved ASR performance compared to previous approaches.
\par

\noindent \textbf{Speech Emotion Recognition}: \citet{chen2023exploring} leveraged Wav2vec2 representations for SER followed by the use of WavLM by \citet{diatlova2024adapting}. Building upon this, \citet{chetiaphukan23_interspeech} presented a unique view, where they showed SFM primarily trained for ASR provides better representations than other SFMs for SER. 

\par

\noindent \textbf{Gender Recognition}: \citet{lebourdais22_interspeech} 
used SOTA PTM wavLM representations for GR as a additional task together with overlapped speech detection. \citet{lee23g_interspeech} has leveraged HuBERT and modeled GR as a auxiliary task together with language indetification for improved SER.

\noindent \textbf{Age Estimation}: \citet{9979878} presented an investigative study into various self-supervised SFMs such as Wav2vec2, WavLM, etc for AE. Further, \citet{gupta22_interspeech} proposed a bi-encoder based mixture of experts model coupled with Wav2vec2 for jointly modeling AE and height estimation. \par

From these studies, we observe that initial efforts have been made to model SFTs simultaneously with SFMs; however, there has been no comprehensive study investigating various SOTA SFMs in this context. Therefore, in this study, we aim to address this gap.

\section{Speech Foundation Models}
\label{section:first_section} 

In this section, we briefly describe the SOTA SFMs utilized in our study.

\noindent \textbf{XLS-R \cite{babu22_interspeech}}: It is a multilingual representation learning model based on Wav2vec2 architecture, trained on 436k hours of speech data and 128 languages. 
We utilize the base version comprising 1 billion parameters\footnote{\url{https://huggingface.co/facebook/wav2vec2-xls-r-1b}}.

\noindent \textbf{Whisper \cite{radford2023robust}}: It is encoder-decoder model pre-trained on 680k hours of data using a multitask format in a weakly-supervised way. Whisper excels in speech recognition, outperforming XLS-R, and we employ the base version with 74 million parameters\footnote{\url{https://huggingface.co/openai/whisper-base}}.

\noindent \textbf{Wav2vec2\cite{baevski2020wav2vec}}: It is pre-trained on 960 hours of Librispeech, and it performs a contrastive task on masked latent representations during pre-training. We use the base version with 95.04 million parameters\footnote{\url{https://huggingface.co/facebook/wav2vec2-base}}.

\noindent \textbf{Massively Multilingual Speech (MMS)\cite{pratap2024scaling}}: Built on the Wav2vec2 architecture and pre-trained on over 500k hours of speech data, MMS uses contrastive pre-training similar to Wav2vec2. We use the 1 billion parameters version\footnote{\url{https://huggingface.co/facebook/mms-1b}}.

\noindent \textbf{Unispeech-SAT\cite{chen2022unispeech}}: It is a contrastive loss model utilizing multitask learning with speaker-aware pre-training. It is pre-trained on 960 hours of Librispeech English data. We utilize the base version with 94.68 million parameters\footnote{\url{https://huggingface.co/microsoft/unispeech-sat-base}}.

\noindent \textbf{WavLM\cite{chen2022wavlm}}: It is a SOTA SFM on SUPERB trained for general-purpose speech representation learning. We use the base version\footnote{\url{https://huggingface.co/microsoft/wavlm-base}} of 94.70 million parameters. 

\noindent \textbf{x-vector\cite{snyder2018x}\footnote{\url{https://huggingface.co/speechbrain/spkrec-xvect-voxceleb}}}: It is a time-delay network trained in a supervised manner for speaker recognition, trained on Voxceleb1 and VoxCeleb2 datasets. It consists of approximately 4.2 million parameters.

\noindent \textbf{TRILLsson \cite{shor22_interspeech}}: It is developed through teacher-student knowledge distillation from the SOTA CAP12 \cite{shor2022universal}.
 We use 63.4 million parameters version\footnote{\url{https://tfhub.dev/google/nonsemantic-speech-benchmark/trillsson4/1}}.

The audio input is resampled to 16 kHz before being fed into the SFMs. We retrieve representations from the last hidden states of the frozen SFMs by average pooling. We extract representations of 512 (x-vector, Whisper), 768 (WavLM, Wav2vec2, Unispeech-SAT), 1024 (TRILLsson), and 1280 (MMS, XLS-R) - dimension respectively.

\begin{figure}[bt!] 
        \centering
        \includegraphics[width=0.4\textwidth]{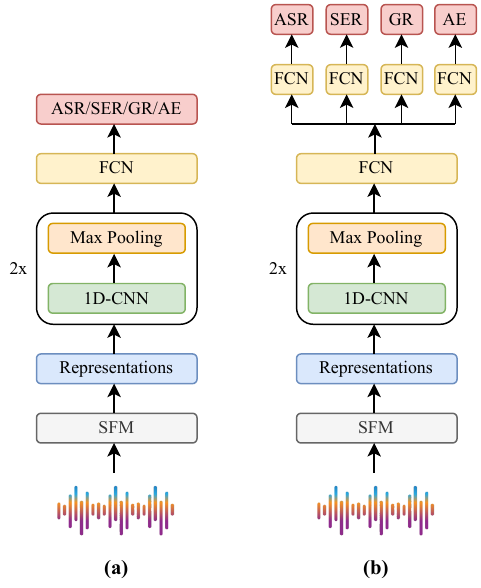} 
        \caption{Single view models: (a) single view single task and (b) single view multi-task.}
        \label{fig:SV}
\end{figure}

\begin{figure*}[hbt!] 
        \centering
        \includegraphics[width=0.99475\textwidth]{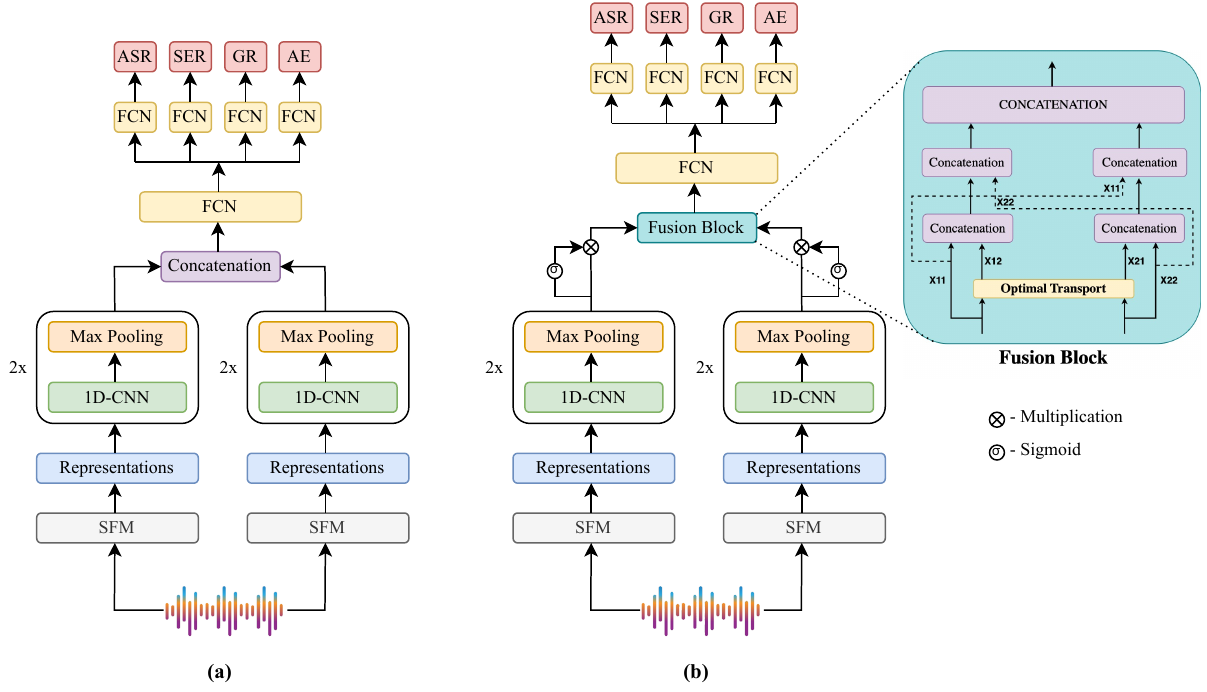} 
        \caption{Multi-view models: (a) Multi View Multi Task with concatenation fusion and (b) \textbf{TANGO}. Here, X11 and X22 denote features from two different views, while X12 and X21 represent features transported from view 2 to the view 1 network and from view 1 to the view 2 network, respectively.}
        \label{fig:MV}
\end{figure*}

\section{Modeling}
In this section, we present various modeling setups for carrying out the experiments in our work. 

\noindent \textbf{Single View Single Task (SVST)}: In SVST, we build downstream models for the SFTs ASR, SER, GR, and AE as individual tasks using different SFM representations.  The architecture is illustrated in Figure \ref{fig:SV} (a). We use two convolution blocks consisting of 1-D CNN and max-pooling layers, followed by a fully connected network (FCN). For output layer, we use the number of neurons depending on the FST.


\noindent \textbf{Single View Multi Task (SVMT)}: We employ individual SFM representations to handle multiple FSTs simultaneously. By integrating task-specific heads after shared layers, this approach facilitates concurrent learning across ASR, SER, GR, and AE. The model architecture followed for SVMT is shown in Figure \ref{fig:SV} (b).

\noindent \textbf{Multi View Multi Task (MVMT)}: It incorporates two SFMs, leveraging their diverse representations to address all four FSTs concurrently. Each view from the SFMs is passed through a network consisting of a convolutional block and FCN, which is the same as in SVST and SVMT. The fusion is performed using concatenation followed by a FCN block. The FCN block is followed by a task-specific head.  The MVMT architecture is presented in Figure \ref{fig:MV} (a).

\noindent The modeling hyperparameters and training details for SVST, SVMT, and MVMT can be found in Appendix \ref{sec:hyper}.


\subsection{TANGO}\label{sec:TANGO}


In this subsection, we present, \textbf{TANGO} for effective MVL for learning multiple SFTs parallely. The modeling architecture diagram is shown in Figure \ref{fig:MV} (b).\textbf{TANGO} also follows the same modeling as SVST, SVMT, and MVMT for each view-specific network, i.e., consisting of two convolutional blocks. After the convolutional blocks, we use a gating mechanism $f(x)$ that consists of a sigmoid function and a multiplication operation. 

\[
f(x) = x \cdot \sigma(x) = x \cdot \frac{1}{1 + e^{-x}}
\]
Here, \( x \) is the input to the function, and \( \sigma(x) \) is the sigmoid function defined as \( \frac{1}{1 + e^{-x}} \). The sigmoid function squashes the input \( x \) into a range between 0 and 1. The resultant value is the scaled output of the original input. This gating mechanism selectively forwards relevant features and enhances task-specific attribute detection. We then attach a fusion block (refer to Figure ~\ref{fig:MV} (b)) that leverages Optimal Transport (OT) or Earth's mover distance as the fusion mechanism. OT measures dissimilarity between views and has shown its effectiveness for multimodal fusion tasks \cite{9706719}. The fusion block will align different views by minimizing their distance during optimization with the OT plan computed using the Sinkhorn algorithm.

The transported features from two SFMs are represented as:
\begin{align}
x_2 \to x_1 &= \gamma \cdot x_2 \\
x_1 \to x_2 &= \gamma^T \cdot x_1
\end{align}
where $x_1$, $x_2$ represents views from two different SFMs and \( \gamma \) is the OT plan derived from the feature distance matrix \( M \):

\begin{equation}
M = \frac{\| x_1 - x_2 \|_2}{\max(\| x_1 - x_2 \|_2)}
\end{equation}

These transported features are then concatenated with the original SFM-specific features to form enriched fused representations. Further, the resultant features are fused again with alternate SFM features through concatenation. The final concatenated features are then passed through an FCN, followed by a task-specific head. \textbf{TANGO} performs joint optimization with task-specific while optimal alignment of different views.

\section{Experiments}

\subsection{Datasets}

\noindent \textbf{CREMA-D} \cite{6849440}: It is a comprehensive resource for ASR, SER, GR, and AE. It offers a rich variety of speaker data, representing diverse ethnicities and age groups, making it an ideal benchmark for training and evaluating ML models. The dataset contains recordings of 7442 words spoken by 91 distinct speakers, balanced by gender, in English. It spans six distinct emotions: anger, happiness, sadness, fear, disgust, and neutrality. The utterances are generated by 43 female and 48 male speakers, each articulating twelve sentences.

\noindent \textbf{BAVED} \cite{BAVED}: It consists of recordings representing three emotional states: Neutral, Low (indicative of tiredness or exhaustion), and High (encompassing emotions such as happiness, joy, sadness, and anger). This dataset comprises a total of 1935 recordings contributed by 45 male and 16 female speakers, all in Arabic.

\noindent \textbf{emo-DB} \cite{burkhardt05b_interspeech}: It contains 535 speech samples recorded from ten actors, with equal representation of genders. Each actor performed ten varied scripts tailored for everyday communication, ensuring a wide range of emotional expression within the dataset.

\begin{figure}[bt] 
        \centering
        \includegraphics[width=0.475\textwidth]{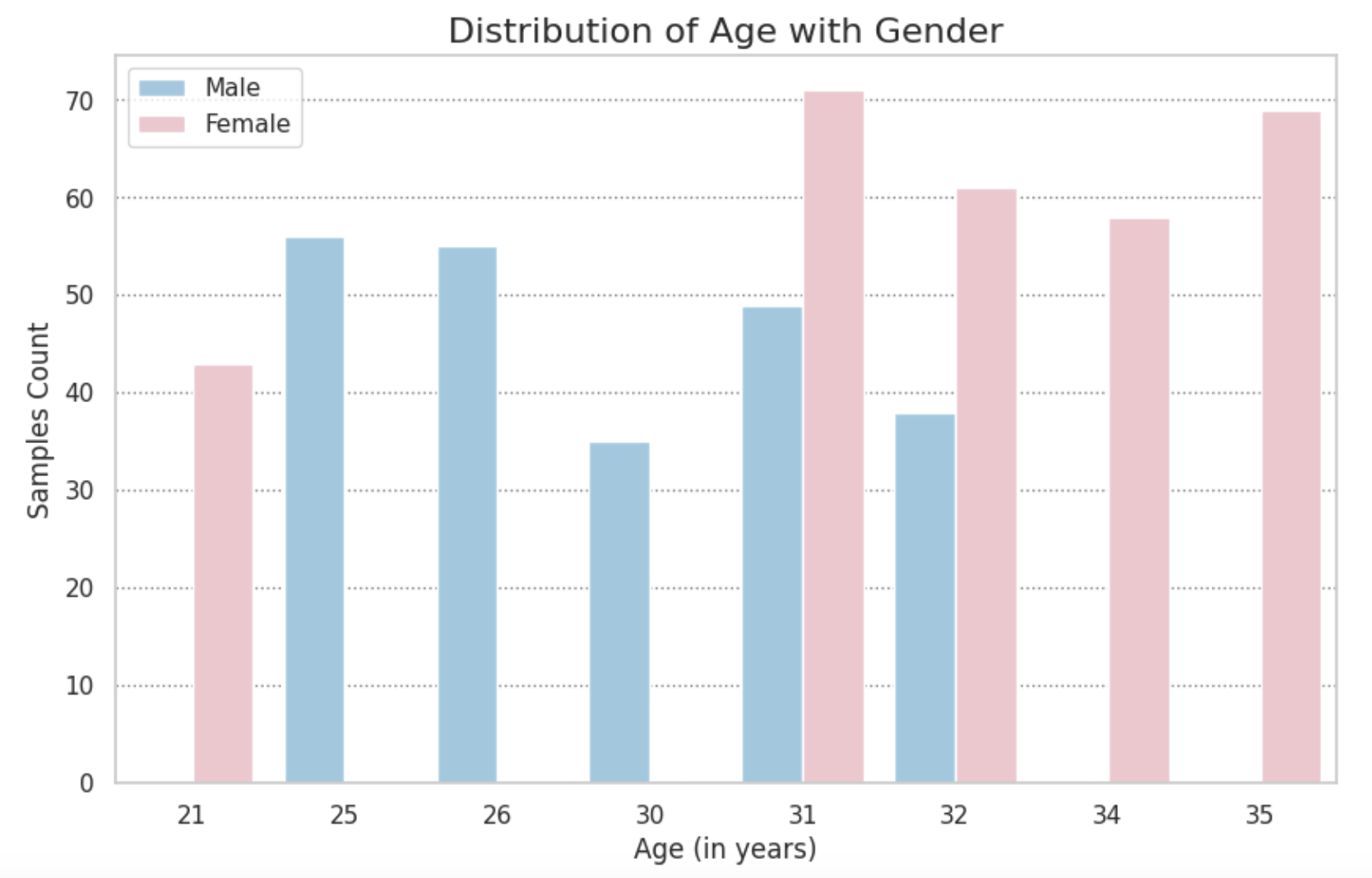} 
        \caption{Age and Gender Distribution for emo-DB.}
        \label{fig:emodbageemo}
\end{figure}

\noindent Further details regarding the datasets gender and age distribution can be found in Appendix Figure \ref{fig:cremadageemo}, \ref{fig:bavedageemo} and \ref{fig:emodbageemo}. 

\subsection{Loss Function}
The multi-task total loss can be expressed as:
\[
\mathcal{L}_{\text{total}} = \lambda_1 \mathcal{L}_{\text{CE1}} + \lambda_2 \mathcal{L}_{\text{CE2}} + \lambda_3 \mathcal{L}_{\text{BCE}} + \lambda_4 \mathcal{L}_{\text{RMSE}}
\]

\noindent where: ${L}_{\text{total}}$, ${L}_{\text{CE1}}$, ${L}_{\text{CE2}}$, ${L}_{\text{BCE}}$, ${L}_{\text{RMSE}}$ represents total loss, cross-entropy (ASR), cross-entropy (SER), BCE (GR), RMSE (AE) respectively. $\lambda_1$, $\lambda_2$, $\lambda_3$, $\lambda_4$ represents weight parameter for ASR, SER, GR, AE losses respectively. In our experiments for multi-task learning, we kept the weightage same as 0.33.

\begin{table}[bt]\label{table1}
\scriptsize
  \centering
  \begin{tabular}{l|c|c|c|c}
    \toprule
    \textbf{SFMs} & \textbf{ASR} & \textbf{SER} & \textbf{GR} & \textbf{AE} \\
    \midrule
    \multicolumn{5}{c}{\textbf{\textit{CREMA-D}}} \\
    \midrule
    XLS-R  & 63.78 & 73.07 & 96.78 & 12.37 \\
    MMS  & 73.54 & 76.76 & 98.66 & 15.45 \\
    Whisper  & 51.98 & 73.20 & 98.25 & 11.14 \\
    Unispeech-SAT & 61.72 & 71.18 & 98.88 & 8.55 \\
    MFCC & 82.04 & 48.56 & 94.96 & 18.02 \\
    Wav2Vec2 & 60.44 & 63.90 & 98.05 & 11.27 \\
    x-vector & 87.51 & 68.25 & 99.23 & 8.80 \\
    \textbf{TRILLsson} & \cellcolor{blue!25}\textbf{90.93} & \cellcolor{blue!25}\textbf{80.15} & \cellcolor{blue!25}\textbf{99.57} & \cellcolor{blue!25}\textbf{5.81} \\
    WavLM & 69.84 & 73.00 & 98.79 & 10.46 \\
    \midrule
    \multicolumn{5}{c}{\textbf{\textit{BAVED}}} \\
    \midrule
    XLS-R  & 87.86 &85.79  &98.71  & 4.07 \\
    MMS  & 81.40 &86.56  &98.71  & 7.72 \\
    Whisper  &80.88  &82.69  &99.74  &3.73  \\
    Unispeech-SAT &83.98  &81.91  &99.22  &6.74  \\
    MFCC & 81.14 &84.75  & 99.48 &5.06  \\
    Wav2Vec2 &82.43  &77.00 &99.74  &3.71  \\
    \textbf{x-vector} &83.20  &87.34  &\cellcolor{blue!25}\textbf{100.00} &3.20   \\
    \textbf{TRILLsson} &\cellcolor{blue!25} \textbf{90.19} &\cellcolor{blue!25} \textbf{88.11} & \cellcolor{blue!25}\textbf{100.00} &\cellcolor{blue!25} \textbf{2.34} \\
    WavLM &87.08  & 82.17 &99.22  &3.94  \\
    \midrule
    \multicolumn{5}{c}{\textbf{\textit{emo-DB}}} \\
    \midrule
    XLS-R  &41.12  & 92.52 & 95.33 & 3.27 \\
    MMS  & 47.66 & 75.70 & 98.13& 4.57 \\
    \textbf{Whisper}  &66.36  &85.98  & \cellcolor{blue!25} \textbf{100.00}  &8.72  \\
    Unispeech-SAT &84.11  &96.26  &95.33  &4.61  \\
    \textbf{MFCC} &75.70  & 69.16 &\cellcolor{blue!25} \textbf{100.00}&2.19 \\
    Wav2Vec2 &75.70  &92.52  & 99.07 &11.10  \\
    \textbf{x-vector} &\cellcolor{blue!25} \textbf{100.00}  &93.46 & \cellcolor{blue!25} \textbf{100.00} & 2.03 \\
    \textbf{TRILLsson} & 98.13 &  \cellcolor{blue!25} \textbf{97.20} &  \cellcolor{blue!25} \textbf{100.00} &  \cellcolor{blue!25} \textbf{1.74} \\
    WavLM &80.37  & 91.59  & \cellcolor{blue!25} \textbf{100.00} &2.11  \\
    \bottomrule
  \end{tabular}
  \caption{Accuracy (ASR, SER, GR) in percentage and RMSE (AE) values for different SFMs in Single View Single Task setup.}
  \label{table:svst}
\end{table}

\begin{table}[bt!]
\scriptsize
  \centering
  \begin{tabular}{l|c|c|c|c}
    \toprule
    \textbf{SFMs} & \textbf{ASR} & \textbf{SER} & \textbf{GR} & \textbf{AE} \\
    \midrule
    \multicolumn{5}{c}{\textbf{\textit{CREMA-D}}} \\
    \midrule
    XLS-R  & 40.73 & 46.41 & 90.43 & 10.79\\
    MMS  & 27.54 & 36.56 & 92.14 & 25.69 \\
    Whisper  & 27.74 & 48.15 & 93.49 & 20.63  \\
    Unispeech-SAT & 52.78 & 61.58 & 97.38 & 9.83 \\
    MFCC &  80.89 & 36.13 & 92.48 & 8.84 \\
    Wav2Vec2 & 53.59  & 55.10 & 96.44 & 10.22 \\
    x-vector & 77.77 & 56.08 & 97.72 & 7.59 \\
    \textbf{TRILLsson} & \cellcolor{blue!25} \textbf{81.13} & \cellcolor{blue!25} \textbf{75.22} & \cellcolor{blue!25} \textbf{98.79} & \cellcolor{blue!25} \textbf{6.87} \\
    WavLM & 58.43 & 63.26  & 97.68 & 9.14 \\
    \midrule
    \multicolumn{5}{c}{\textbf{\textit{BAVED}}} \\
    \midrule
    XLS-R  & 79.54 & 66.79 & 94.18 & 4.99 \\
    MMS  & 67.70 & 45.99 & 86.43 & 7.03 \\
    Whisper  & 57.11 & 68.99 & 78.04 & 9.44 \\
    Unispeech-SAT & 83.98 & 68.22 & 98.71 & 3.30 \\
    MFCC & 87.34 & 69.51 & 98.45 & 2.84 \\
    Wav2Vec2 & 81.40 & 70.80 & 98.71 & 5.87 \\
    x-vector & 84.50 &  81.40 & 99.48 & 2.56 \\
    \textbf{TRILLsson} & \cellcolor{blue!25} \textbf{89.41} & \cellcolor{blue!25}\textbf{83.98} & \cellcolor{blue!25}\textbf{100.00} & \cellcolor{blue!25} \textbf{2.58} \\
    WavLM & 83.72 & 72.09 & 98.71 & 3.52 \\
    \midrule
    \multicolumn{5}{c}{\textbf{\textit{emo-DB}}} \\
    \midrule
    XLS-R  & 61.68 & 74.77 & 92.52 & 3.48 \\
    MMS  & 58.88 & 69.16 & 72.43 & 5.93 \\
    Whisper  & 46.73 & 62.62 & 90.19 & 10.94 \\
    Unispeech-SAT & 81.31 & 86.92 & 95.79 & 4.25 \\
    MFCC & 85.98 & 66.35 & 96.26 & 2.12 \\
    Wav2Vec2 & 74.77 & 81.31 & 97.20 &  2.63 \\
    x-vector & 95.33 & 87.85 & 98.13 & 3.57 \\
    \textbf{TRILLsson} & \cellcolor{blue!25} \textbf{96.26} & \cellcolor{blue!25} \textbf{95.33} & \cellcolor{blue!25} \textbf{99.07} & \cellcolor{blue!25} \textbf{2.00} \\
    WavLM & 88.31 & 82.71 & 96.27 & 2.65 \\
    \bottomrule
  \end{tabular}
  \caption{Accuracy (ASR, SER, GR) in percentage and RMSE (AE) metrics for various SFMs in Single View Multi Task setup.}
  \label{table:svmt}
\end{table}

\begin{table}[hbt!]
\scriptsize
  \centering
  \begin{tabular}{l|c|c|c|c}
    \toprule
    \textbf{SFMs} & \textbf{ASR} & \textbf{SER} & \textbf{GR} & \textbf{AE} \\
    \midrule
    \multicolumn{5}{c}{\textbf{\textit{Concatenation Fusion Technique}}} \\
    \midrule
    XLS-R + MMS & 21.96 & 33.24 & 95.63 & 18.07\\
    XLS-R + Whisper & 51.18 & 38.01 & 94.49 & 12.38 \\
    XLS-R + Unispeech-SAT & 61.72 & 60.11 & 97.38 & 11.54 \\
    XLS-R + MFCC & 78.11 & 42.91 & 96.71 & 8.20 \\
    XLS-R + Wav2vec2 & 61.05 & 52.92 & 91.87 & 9.57 \\
    XLS-R + x-vector & 78.91 & 60.51 & 97.99 & 7.13 \\
    XLS-R + TRILLsson & 80.39 & 72.33 & 99.19 & 6.64 \\
    XLS-R + WavLM & 62.79 & 63.26 & 98.05 & 9.25 \\
    MMS + Whisper & 29.75 & 40.23 & 98.46 & 10.36 \\
    MMS + Unispeech-SAT & 51.18 & 60.04 & 97.25 & 9.38 \\
    MMS + MFCC & 39.89 & 37.07 & 97.31 & 9.79 \\
    MMS + Wav2Vec2 & 50.57 & 55.34 & 97.11 & 8.97 \\
    MMS + x-vector & 63.40 & 55.94 & 97.78 & 8.05 \\
    MMS + TRILLsson & 80.59 & 71.73 & 99.13 & 7.11 \\
    MMS + WavLM & 58.83 & 54.53 & 99.06 & 8.25 \\
    Whisper + Unispeech-SAT & 42.04 & 58.16 & 97.38 & 9.24 \\
    Whisper + MFCC & 66.49 & 35.33 & 96.91 & 9.79 \\
    Whisper + Wav2Vec2 & 31.50 & 53.26 & 92.48 & 10.77 \\
    Whisper + x-vector & 76.02 & 60.91 & 98.12 & 7.55 \\
    Whisper + TRILLsson & 81.60 & 74.21 & 99.40 &  6.74\\
    Whisper + WavLM & 48.62 & 60.24 & 97.85 & 9.74 \\
    Unispeech-SAT + MFCC & 79.85 & 63.60 & 97.58 & 9.03 \\
    Unispeech-SAT + Wav2Vec2 & 64.00 & 60.85 & 98.39 & 8.59  \\
    Unispeech-SAT + x-vector & 80.32 & 66.22 & 98.79 & 6.99 \\
    Unispeech-SAT + TRILLsson & 83.21 & 75.22 & 99.19 & 6.31 \\
    Unispeech-SAT + WavLM & 63.00 & 66.42 & 98.93 & 8.81 \\
    MFCC + Wav2Vec2 & 82.94 & 50.77 & 97.38 & 8.20 \\
    MFCC + x-vector & 89.59 & 54.87 & 98.79 & 6.42 \\
    MFCC + TRILLsson & 88.18 & 73.81 & 99.46 & 6.32 \\
    MFCC + WavLM & 82.14 & 61.72 & 97.52 & 7.52 \\
    Wav2Vec2 + x-vector & 77.50 & 59.64 & 98.32 & 7.32\\
    Wav2Vec2 + TRILLsson & 83.55 & 75.02 & 99.13 & 6.50 \\
    Wav2Vec2 + WavLM & 68.03 & 64.27 & 98.12 & 8.53 \\
    x-vector + TRILLsson & 87.98 & 74.68 & 99.53 & 5.70 \\
    x-vector + WavLM & 78.17 & 62.39 & 62.39 & 7.89 \\
    TRILLsson + WavLM & 83.81 & 73.61 & 98.79 & 6.69 \\

    \midrule
    \multicolumn{5}{c}{\textbf{\textit{TANGO}}} \\
    \midrule

    XLS-R + MMS & 27.20 & 38.82 & 80.12  & 10.97 \\
    XLS-R + Whisper & 38.42 & 54.37 & 93.42 & 9.38 \\
    XLS-R + Unispeech-SAT & 57.59 & 59.90 & 96.78 & 9.01 \\
    XLS-R + MFCC & 65.21 & 51.91 & 95.97 & 9.38 \\
    XLS-R + Wav2vec2 & 55.30 & 60.11 & 97.11 & 10.40 \\
    XLS-R + x-vector & 77.84 & 60.91 & 96.87 & 9.44 \\
    XLS-R + TRILLsson & 84.15 & 73.67 & 99.23 & 6.58 \\
    XLS-R + WavLM & 59.10 & 65.48 & 91.78 & 9.50 \\
    MMS + Whisper & 29.21 & 45.26 & 97.81 & 11.83 \\
    MMS + Unispeech-SAT & 38.24 & 56.68 & 91.40 & 11.10 \\
    MMS + MFCC & 49.56 & 53.69 & 96.17 & 10.74 \\
    MMS + Wav2Vec2 & 33.11 & 51.44 & 92.48 & 15.64 \\
    MMS + x-vector & 57.15 & 57.89 & 97.21 & 11.57 \\
    MMS + TRILLsson & 80.32 & 69.30 & 99.06 & 6.72 \\
    MMS + WavLM & 39.25 & 35.46 & 93.65 & 14.60 \\
    Whisper + Unispeech-SAT & 49.26 & 57.99 & 97.85 & 9.88 \\
    Whisper + MFCC &  68.67 & 57.99 & 95.56 & 10.60 \\
    Whisper + Wav2Vec2 & 45.30 & 52.18 & 92.08 & 8.88 \\
    Whisper + x-vector & 75.18 & 62.66 & 98.76 & 7.36 \\
    Whisper + TRILLsson & 81.80 & 74.54 & 99.33 & 6.56 \\
    Whisper + WavLM & 54.10 & 62.29 & 97.92 & 10.02 \\
    Unispeech-SAT + MFCC & 82.54 & 63.47 & 98.19 & 6.91 \\
    Unispeech-SAT + Wav2Vec2 & 63.06 & 61.35 & 97.65 & 9.11 \\
    Unispeech-SAT + x-vector & 83.61 & 65.14 & 98.99 &  7.13\\
    Unispeech-SAT + TRILLsson & 83.41 & 75.18 & 99.26 & 6.43 \\
    Unispeech-SAT + WavLM & 64.71 & 65.54 & 98.39 & 8.61 \\
    MFCC + Wav2Vec2 & 82.74 & 54.60 & 90.97 & 7.75 \\
    MFCC + x-vector & 90.01 & 61.25 & 97.78 & 5.98 \\
    MFCC + TRILLsson & 88.99 & 74.98 & 99.19 & 5.85 \\
    MFCC + WavLM & 84.62 & 65.28 & 97.92 & 7.58 \\
    Wav2Vec2 + x-vector & 82.84 & 61.18 & 98.72 & 7.04 \\
    Wav2Vec2 + TRILLsson & 83.34 & 75.02 & 98.86 & 6.45 \\
    Wav2Vec2 + WavLM & 66.22 & 63.73 & 97.85 &  9.09 \\
    \textbf{x-vector + TRILLsson} & \cellcolor{blue!25} \textbf{90.19} & \cellcolor{blue!25} \textbf{75.85} & \cellcolor{blue!25} \textbf{99.60} & \cellcolor{blue!25} \textbf{5.68} \\
    x-vector + WavLM & 84.11 & 68.00 & 98.59 & 6.50 \\
    TRILLsson + WavLM & 84.89 & 75.55 & 99.46 & 6.40 \\
    \bottomrule
  \end{tabular}
  \caption{Accuracy (ASR, SER, GR) in percentage and RMSE (AE) values for different SFMs in Multi-View Multi-Task setup for CREMA-D.}
  \label{table:mvmtc}
\end{table}

\begin{table}[hbt!]
\scriptsize
  \centering
  \begin{tabular}{l|c|c|c|c}
    \toprule
    \textbf{SFMs} & \textbf{ASR} & \textbf{SER} & \textbf{GR} & \textbf{AE} \\
    \midrule
    \multicolumn{5}{c}{\textbf{\textit{Concatenation Fusion Technique}}} \\
    \midrule
         XLS-R + MMS & 83.98 & 74.16 & 97.67 & 3.50\\
    XLS-R + Whisper & 84.50 & 77.00 & 98.71 & 3.76 \\
    XLS-R + Unispeech-SAT & 85.53 & 80.88 & 98.71 & 3.14 \\
    XLS-R + MFCC & 87.34 & 83.72 & 98.45 & 2.89 \\
    XLS-R + Wav2vec2 & 84.75 & 81.40 & 98.45 & 3.63 \\
    XLS-R + x-vector & 79.84 & 83.98 & 99.22 & 2.59 \\
    \textbf{XLS-R + TRILLsson} & 88.63 & 85.01 & \cellcolor{blue!25} \textbf{100.00} & 2.12 \\
    XLS-R + WavLM & 86.05 & 78.81 & 97.16 & 3.43 \\
    MMS + Whisper & 79.33 & 81.91 &96.90  &  3.71 \\
    MMS + Unispeech-SAT & 85.01 & 79.33 & 97.93 & 4.04 \\
    MMS + MFCC & 81.40 & 74.68 & 98.45 &3.74  \\
    MMS + Wav2Vec2 & 80.62 & 79.59 & 97.16 & 3.82 \\
    \textbf{MMS + x-vector} & 84.75 & 83.72 & \cellcolor{blue!25} \textbf{100.00} & 2.84 \\
    MMS + TRILLsson & 86.05 & 87.08 & 99.74 & 1.69 \\
    MMS + WavLM & 85.79 & 78.04 & 98.45 & 3.70 \\
    Whisper + Unispeech-SAT & 81.65 & 75.71 & 99.74 & 3.22 \\
    Whisper + MFCC & 87.08 & 75.45 & 98.71 & 3.13 \\
    Whisper + Wav2Vec2 & 80.88 & 80.88 & 99.74 & 3.46 \\
    Whisper + x-vector & 81.40 & 82.43 & 99.48 & 3.94 \\
    \textbf{Whisper + TRILLsson} & 87.60 & 87.08 & \cellcolor{blue!25} \textbf{100.00} & 2.15 \\
    Whisper + WavLM & 83.20 & 80.36 & 98.97 & 2.94 \\
    Unispeech-SAT + MFCC & 88.11 & 80.10 & 99.48 & 2.39 \\
    Unispeech-SAT + Wav2Vec2 & 86.30 & 78.55 & 98.71 & 3.04 \\
    Unispeech-SAT + x-vector & 87.08 & 81.65 & 98.97 & 2.78 \\
    \textbf{Unispeech-SAT + TRILLsson} & 89.41 & 83.20 & \cellcolor{blue!25} \textbf{100.00}  & 2.12  \\
    Unispeech-SAT + WavLM & 87.86 & 79.33 & 98.19 & 3.22 \\
    MFCC + Wav2Vec2 & 90.18 & 78.29 & 99.22 & 2.79 \\
    MFCC + x-vector & 89.15 & 82.69 & 99.48 & 1.77 \\
    MFCC + TRILLsson & 91.21 & 86.82 & 99.74 & 2.17 \\
    MFCC + WavLM & 88.63 &  82.69 &  99.48 & 3.04 \\
    Wav2Vec2 + x-vector & 89.41 & 81.91 & 99.74 & 2.22 \\
    \textbf{Wav2Vec2 + TRILLsson} & 89.15 & 85.27 & \cellcolor{blue!25} \textbf{100.00} & 1.92 \\
    Wav2Vec2 + WavLM & 87.86 & 82.17 & 99.22 & 3.11 \\
    x-vector + TRILLsson & 89.15 & 86.56 & 99.74 & 1.73 \\
    x-vector + WavLM & 89.92 & 82.69 & 99.72 & 2.10 \\
    TRILLsson + WavLM & 90.44 & 86.05 & 99.74 & 1.76 \\
    \midrule
    \multicolumn{5}{c}{\textbf{\textit{TANGO}}} \\
    \midrule
        XLS-R + MMS & 84.36 & 73.90 & 98.45 & 3.62 \\
    XLS-R + Whisper & 72.09 & 69.38 & 98.71 & 3.50 \\
    XLS-R + Unispeech-SAT & 85.53 & 79.84 & 99.48 & 3.02 \\
    XLS-R + MFCC & 90.69 & 72.87 & 97.67 & 3.37 \\
    XLS-R + Wav2vec2 & 85.65 & 74.16 & 99.48 & 2.90 \\
    \textbf{XLS-R + x-vector} & 88.11 & 75.06 & \cellcolor{blue!25} \textbf{100.00} & 2.67 \\
    XLS-R + TRILLsson & 90.18 & 85.01  & 99.87 & 1.59 \\
    XLS-R + WavLM & 82.68 & 74.54 & 98.97 & 3.17 \\
    MMS + Whisper & 76.35 & 58.65 & 96.51 & 3.88 \\
    MMS + Unispeech-SAT & 86.82 & 75.45 & 98.45 & 3.18 \\
    MMS + MFCC & 77.39 & 79.58 & 98.19 & 4.03 \\
    MMS + Wav2Vec2 & 79.20 & 78.55 & 99.09 & 3.54 \\
    MMS + x-vector & 83.20 & 75.19 & 98.71 & 3.85 \\
    MMS + TRILLsson & 90.44 & 85.79 & 99.87 & 1.79 \\
    MMS + WavLM & 86.05 & 72.61 & 99.61 & 3.33 \\
    Whisper + Unispeech-SAT & 83.46 & 75.19 & 99.74 & 2.82 \\
    Whisper + MFCC & 83.07 & 69.51 & 98.45 & 3.94 \\
    Whisper + Wav2Vec2 & 79.46 & 71.83 & 98.71 & 4.38  \\
    Whisper + x-vector & 84.50 & 82.43 & 98.45 & 4.39 \\
    Whisper + TRILLsson & 90.18 & 85.01 & 99.87 & 2.14 \\
    Whisper + WavLM & 83.85 & 75.84 & 99.74 & 3.15 \\
    Unispeech-SAT + MFCC & 88.24 & 72.22 & 99.48 & 2.26 \\
    Unispeech-SAT + Wav2Vec2 & 87.08 & 77.00 & 98.71 & 2.91 \\
    Unispeech-SAT + x-vector & 86.43 & 77.39 & 99.22 & 1.94 \\
    \textbf{Unispeech-SAT + TRILLsson} & 90.44 & 87.08 & \cellcolor{blue!25} \textbf{100.00} & 1.90 \\
    Unispeech-SAT + WavLM & 83.20 & 77.51 & 97.67 & 3.16 \\
    MFCC + Wav2Vec2 & 88.89 & 75.84 & 98.71 & 2.39 \\
    MFCC + x-vector & 91.47 & 78.68 & 99.74 & 2.25 \\
    MFCC + TRILLsson & 90.96 & 86.30 & 99.87 & 2.25 \\
    MFCC + WavLM & 87.85 & 79.20 & 99.74 &  2.25 \\
    Wav2Vec2 + x-vector & 87.46 & 79.19 & 99.48 & 1.75 \\
    \textbf{Wav2Vec2 + TRILLsson} & 91.47 & 84.50 & \cellcolor{blue!25} \textbf{100.00} & 1.74 \\
    Wav2Vec2 + WavLM & 85.91 & 75.84 & 98.97 & 5.39 \\
    \textbf{x-vector + TRILLsson} & \cellcolor{blue!25} \textbf{91.99} & \cellcolor{blue!25} \textbf{87.60} & \cellcolor{blue!25} \textbf{100.00} & \cellcolor{blue!25} \textbf{1.44} \\
    x-vector + WavLM & 90.05 & 78.16 & 99.74 & 3.85 \\
    \textbf{TRILLsson + WavLM} & 91.08 & 83.98 & \cellcolor{blue!25} \textbf{100.00} & 1.58 \\
    \bottomrule
  \end{tabular}
  \caption{Accuracy (ASR, SER, GR) in percentage and RMSE (AE) values for different SFMs in Multi-View Multi-Task Setup for BAVED.}
  \label{table:mvmtb}
\end{table}

\begin{table}[hbt!]
\scriptsize
  \centering
  \begin{tabular}{l|c|c|c|c}
    \toprule
    \textbf{SFMs} & \textbf{ASR} & \textbf{SER} & \textbf{GR} & \textbf{AE} \\
    \midrule
    \multicolumn{5}{c}{\textbf{\textit{Concatenation Fusion Technique}}} \\
    \midrule
    XLS-R + MMS & 51.40 & 77.10 & 81.31 & 2.77  \\
    XLS-R + Whisper & 61.21 & 77.57 & 98.13 & 4.67\\
    XLS-R + Unispeech-SAT & 82.24 &  85.51 & 98.13 & 2.46 \\
    XLS-R + MFCC & 91.12 & 76.17 & 97.20 & 3.23 \\
    XLS-R + Wav2vec2 & 80.84 & 85.98 & 93.46 & 3.27 \\
    \textbf{XLS-R + x-vector} & 97.20 & 83.64 & \cellcolor{blue!25} \textbf{100.00} & 2.06 \\
    \textbf{XLS-R + TRILLsson} & 97.86 & 96.73 & \cellcolor{blue!25} \textbf{100.00} & 1.49 \\
    XLS-R + WavLM & 83.18 & 87.38 & 99.53 & 2.82 \\
    MMS + Whisper & 63.08 & 71.96 & 94.39 & 2.73 \\
    MMS + Unispeech-SAT & 57.47 & 85.04 & 91.59 & 2.27 \\
    MMS + MFCC & 85.51 & 71.96 & 98.13 & 2.03 \\
    MMS + Wav2Vec2 & 69.16 & 75.23 & 97.66 & 3.73 \\
    MMS + x-vector & 97.19 & 87.85 & 97.66 & 1.45 \\
    MMS + TRILLsson & 98.13 & 95.79 & 99.07 & 1.93 \\
    MMS + WavLM & 78.50 & 85.05 & 96.73 & 2.16 \\
    Whisper + Unispeech-SAT & 71.49 & 81.31 & 98.13 & 2.53 \\
    Whisper + MFCC & 77.10 & 85.51 & 93.93 & 2.31 \\
    Whisper + Wav2Vec2 & 74.30 & 81.30 & 95.79 & 2.75\\
    Whisper + x-vector & 97.66 & 86.91 & 99.07 & 1.82 \\
    Whisper + TRILLsson & 93.92 & 92.99 & 99.53 & 1.98 \\
    Whisper + WavLM & 85.51 & 85.04 & 93.46  & 2.56 \\
    Unispeech-SAT + MFCC & 90.18 & 85.98 & 95.33 & 2.09 \\
    Unispeech-SAT + Wav2Vec2 & 87.85 & 88.78 & 97.66 & 2.66 \\
    \textbf{Unispeech-SAT + x-vector} & 97.20 & 83.64 & \cellcolor{blue!25} \textbf{100.00} & 2.06 \\
    \textbf{Unispeech-SAT + TRILLsson} & 96.73 & 99.07 & \cellcolor{blue!25} \textbf{100.00} & 3.61 \\
    Unispeech-SAT + WavLM & 86.91 & 87.85 & 99.06 & 2.18 \\
    MFCC + Wav2Vec2 & 94.39 & 86.91 & 99.07 & 2.17 \\
    MFCC + x-vector & 97.19 & 86.44 & 99.07 & 3.43 \\
    MFCC + TRILLsson & 96.26 & 91.59 & 99.53 & 2.02 \\
    \textbf{MFCC + WavLM} & 95.33 & 85.98 & \cellcolor{blue!25} \textbf{100.00} & 1.62 \\
    Wav2Vec2 + x-vector & 97.66 & 85.98 & 99.53 & 1.46 \\
    \textbf{Wav2Vec2 + TRILLsson} & 97.19 & 96.26 & \cellcolor{blue!25} \textbf{100.00} & 1.66 \\
    Wav2Vec2 + WavLM & 86.91 & 87.38 & 99.06 & 2.29 \\
    \textbf{x-vector + TRILLsson} & 97.20 & 93.46 & \cellcolor{blue!25} \textbf{100.00} & 1.56 \\
    \textbf{x-vector + WavLM} & 97.20 & 87.85 & \cellcolor{blue!25} \textbf{100.00} & 1.57 \\
    TRILLsson + WavLM & 97.20 & 96.26 & 99.05 & 1.98 \\
    \midrule
    \multicolumn{5}{c}{\textbf{\textit{TANGO}}} \\
    \midrule
    XLS-R + MMS & 71.96 & 72.90 & 95.33 & 2.43 \\
    XLS-R + Whisper & 74.77 & 72.90 & 99.07 & 2.70 \\
    \textbf{XLS-R + Unispeech-SAT} & 81.31 & 95.33 & \cellcolor{blue!25} \textbf{100.00} & 2.52 \\
    XLS-R + MFCC & 90.65 & 79.44 & 98.13 & 2.19 \\
    XLS-R + Wav2vec2 & 80.37 & 84.11 & 99.07  & 2.72 \\
    XLS-R + x-vector & 98.13 & 87.85 & 99.07 & 1.49  \\
    \textbf{XLS-R + TRILLsson} & 96.26 & 98.13 & \cellcolor{blue!25} \textbf{100.00} & 1.65 \\
    \textbf{XLS-R + WavLM} & 90.65 & 91.59 & \cellcolor{blue!25} \textbf{100.00} & 2.15 \\
    MMS + Whisper & 88.79 & 91.59 & 97.20 & 2.14 \\
    \textbf{MMS + Unispeech-SAT} & 79.44 & 88.79 & \cellcolor{blue!25} \textbf{100.00} & 2.45 \\
    MMS + MFCC & 88.79 & 81.31 & 97.20 & 2.21 \\
    MMS + Wav2Vec2 & 86.92 & 86.92 & 98.13 & 2.75 \\
    \textbf{MMS + x-vector} & 96.26 & 89.72 & \cellcolor{blue!25} \textbf{100.00} & 1.83 \\
    \textbf{MMS + TRILLsson} & 99.07 & 99.07 & \cellcolor{blue!25} \textbf{100.00} & 1.87 \\
    MMS + WavLM & 88.79 & 91.59 & 99.07 & 2.18 \\
    \textbf{Whisper + Unispeech-SAT} & 79.44 & 84.11 & \cellcolor{blue!25} \textbf{100.00} & 2.79 \\
    \textbf{Whisper + MFCC} & 87.85 & 83.18 & \cellcolor{blue!25} \textbf{100.00} & 2.45 \\
    Whisper + Wav2Vec2 & 82.24 & 86.92 & 99.07 & 3.71\\
    \textbf{Whisper + x-vector} & 97.20 & 91.59 & \cellcolor{blue!25} \textbf{100.00}  & 1.89 \\
    \textbf{Whisper + TRILLsson} & 95.33 & 97.20  & \cellcolor{blue!25} \textbf{100.00} & 1.99 \\
    \textbf{Whisper + WavLM} & 85.05 & 89.72 & \cellcolor{blue!25} \textbf{100.00} & 2.24 \\
    Unispeech-SAT + MFCC & 89.72 &90.65  & 97.20 &  2.11\\
    Unispeech-SAT + Wav2Vec2 & 91.59 & 90.65 & 99.07 & 2.40 \\
    \textbf{Unispeech-SAT + x-vector} & 94.39 & 88.79 & \cellcolor{blue!25} \textbf{100.00} & 1.74  \\
    \textbf{Unispeech-SAT + TRILLsson} & 96.26 & 99.07 & \cellcolor{blue!25} \textbf{100.00} & 1.68 \\
    \textbf{Unispeech-SAT + WavLM} & 91.59 & 96.26 & \cellcolor{blue!25} \textbf{100.00} & 2.21 \\
    MFCC + Wav2Vec2 & 89.72 & 82.24 & 99.07 & 2.63 \\
    MFCC + x-vector & 81.30 &  87.85& 99.07 & 1.57 \\
    \textbf{MFCC + TRILLsson} & 98.13 & 96.26 & \cellcolor{blue!25} \textbf{100.00} & 1.73 \\
    MFCC + WavLM & 95.33 & 88.79 & 98.13 &  2.31 \\
    \textbf{Wav2Vec2 + x-vector} & 95.33 & 90.65 & \cellcolor{blue!25} \textbf{100.00} &  1.68 \\
    \textbf{Wav2Vec2 + TRILLsson} & 98.13 & 99.07 & \cellcolor{blue!25} \textbf{100.00} & 1.80 \\
    \textbf{Wav2Vec2 + WavLM} & 93.46 & 93.46 & \cellcolor{blue!25} \textbf{100.00} & 2.21  \\
   \textbf{x-vector + TRILLsson} & \cellcolor{blue!25} \textbf{99.53} & \cellcolor{blue!25} \textbf{100.00} & \cellcolor{blue!25} \textbf{100.00} & \cellcolor{blue!25} \textbf{1.38}\\
    \textbf{x-vector + WavLM} & 96.26 & 95.33 & \cellcolor{blue!25} \textbf{100.00} & 1.70 \\
    \textbf{TRILLsson + WavLM} & 97.20 & 98.13 & \cellcolor{blue!25} \textbf{100.00} & 1.56  \\
    \bottomrule
  \end{tabular}
  \caption{Accuracy (ASR, SER, GR) in percentage and RMSE (AE) values for different SFMs in Multi-View Multi-Task Setup for emo-DB.}
  \label{table:mvmte}
\end{table}

\begin{table}[bt!]
\scriptsize
  \centering
  \begin{tabular}{l|c|c|c|c}
    \toprule
    \textbf{SFMs} & \textbf{ASR} & \textbf{SER} & \textbf{GR} & \textbf{AE} \\
    \midrule
    \multicolumn{5}{c}{\textbf{\textit{CREMA-D}}} \\
    \midrule
    \textbf{SVST}  & \cellcolor{blue!25}\textbf{90.93} & \cellcolor{blue!25}\textbf{80.15} & \cellcolor{blue!25}\textbf{99.57} & \cellcolor{blue!25}\textbf{5.81}\\
    \textbf{SVMT}  & \cellcolor{blue!25} \textbf{81.13} & \cellcolor{blue!25} \textbf{75.22} & \cellcolor{blue!25} \textbf{98.79} & \cellcolor{blue!25} \textbf{6.87} \\
    \textbf{TANGO}  & \cellcolor{blue!25} \textbf{90.19} & \cellcolor{blue!25} \textbf{75.85} & \cellcolor{blue!25} \textbf{99.60} & \cellcolor{blue!25} \textbf{5.68}  \\
    \midrule
    \multicolumn{5}{c}{\textbf{\textit{BAVED}}} \\
    \midrule
    \textbf{SVST}  & \cellcolor{blue!25} \textbf{90.19} &\cellcolor{blue!25} \textbf{88.11} & \cellcolor{blue!25}\textbf{100.00} &\cellcolor{blue!25} \textbf{2.34}\\
    \textbf{SVMT}  & \cellcolor{blue!25} \textbf{89.41} & \cellcolor{blue!25}\textbf{83.98} & \cellcolor{blue!25}\textbf{100.00} & \cellcolor{blue!25} \textbf{2.58} \\
    \textbf{TANGO}  & \cellcolor{blue!25} \textbf{91.99} & \cellcolor{blue!25} \textbf{87.60} & \cellcolor{blue!25} \textbf{100.00} & \cellcolor{blue!25} \textbf{1.44} \\
    \midrule
    \multicolumn{5}{c}{\textbf{\textit{emo-DB}}} \\
    \midrule
    \textbf{SVST}  &\cellcolor{blue!25} \textbf{100} &\cellcolor{blue!25} \textbf{97.20} & \cellcolor{blue!25}\textbf{100.00} &\cellcolor{blue!25} \textbf{1.74}\\
    \textbf{SVMT}  &\cellcolor{blue!25} \textbf{96.26} & \cellcolor{blue!25} \textbf{95.33} & \cellcolor{blue!25} \textbf{99.07} & \cellcolor{blue!25} \textbf{2.00} \\
    \textbf{TANGO}  & \cellcolor{blue!25} \textbf{99.53} & \cellcolor{blue!25} \textbf{100.00} & \cellcolor{blue!25} \textbf{100.00} & \cellcolor{blue!25} \textbf{1.38}  \\
    \bottomrule
  \end{tabular}
  \caption{Accuracy (ASR, SER, GR) in percentage and RMSE (AE) values for best-performing models across SVST, SVMT, and MVMT with \textbf{TANGO}.}
  \label{table:final_comp}
\end{table}

\subsection{Experimental Results}
\begin{figure*}[htbp]
    \centering
    \begin{subfigure}[b]{0.31\textwidth}
        \centering
        \includegraphics[width=\textwidth]{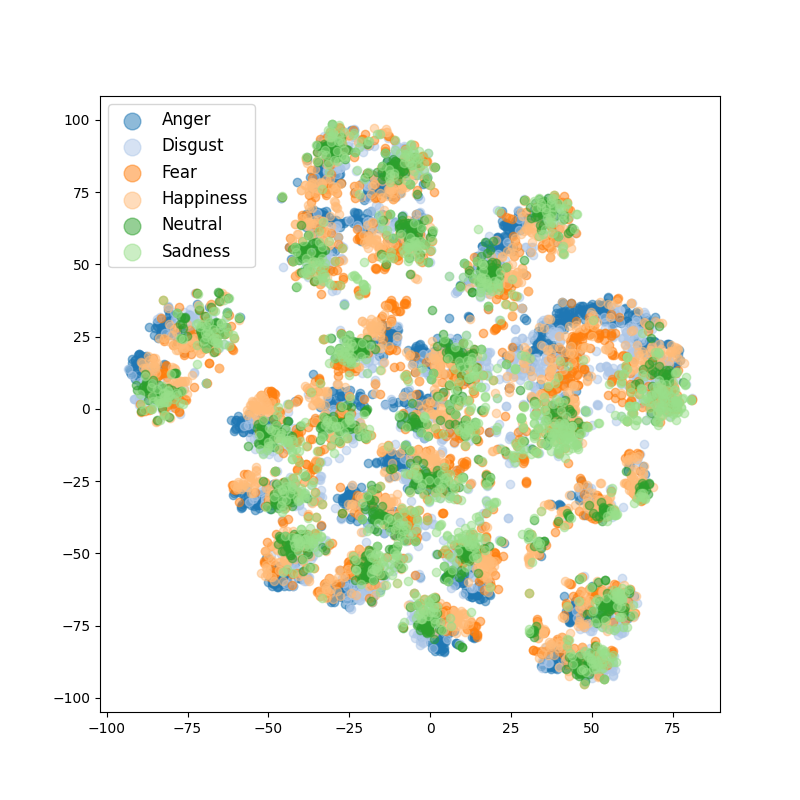}
        \caption{TRILLsson - SER}
    \end{subfigure}
    \hfill
    \begin{subfigure}[b]{0.31\textwidth}
        \centering
        \includegraphics[width=\textwidth]{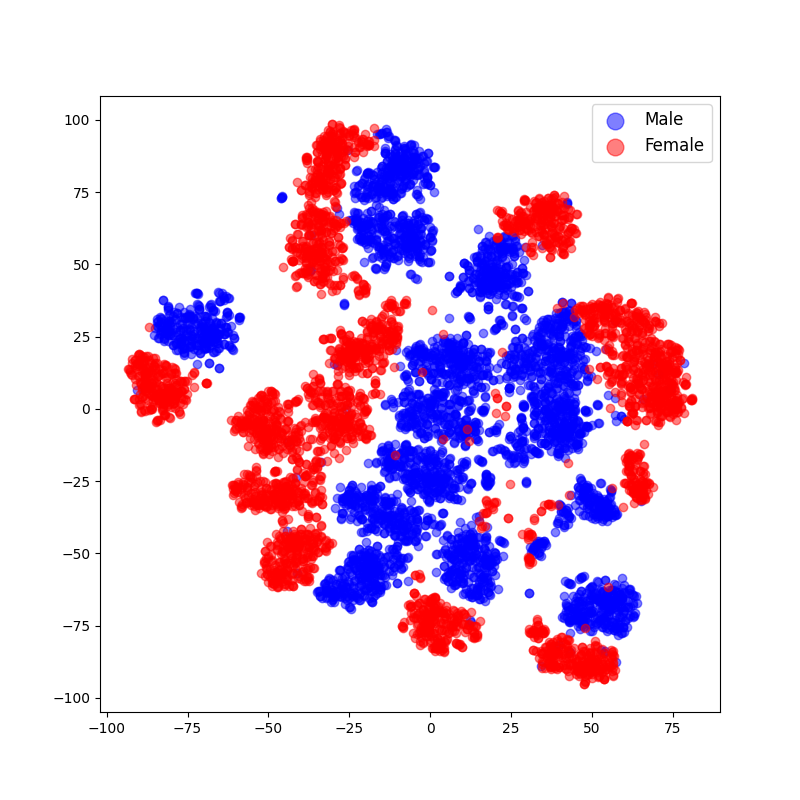}
        \caption{TRILLsson - GR}
    \end{subfigure}
    \hfill
    \begin{subfigure}[b]{0.31\textwidth}
        \centering
        \includegraphics[width=\textwidth]{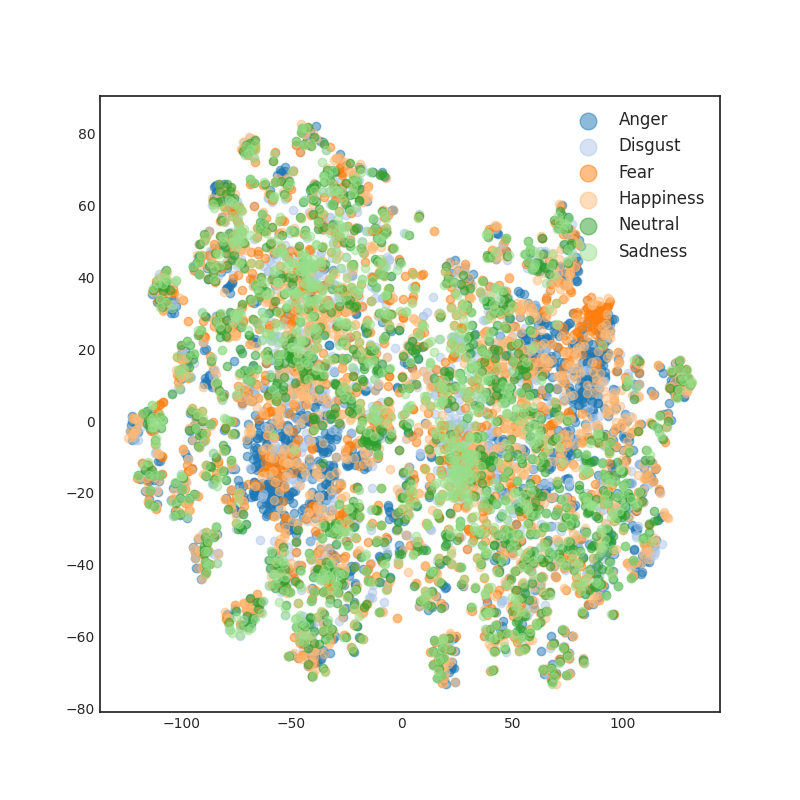}
        \caption{x-vector - SER}
    \end{subfigure}
    
    \vspace{0.5cm} 

    \begin{subfigure}[b]{0.31\textwidth}
        \centering
        \includegraphics[width=\textwidth]{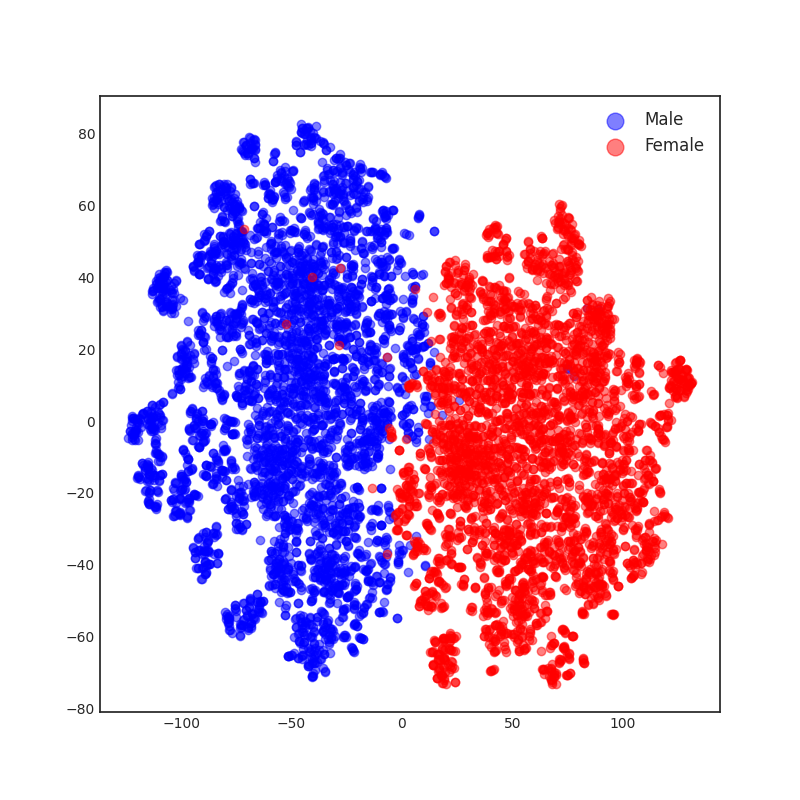}
        \caption{x-vector - GR}
    \end{subfigure}
    \hfill
    \begin{subfigure}[b]{0.31\textwidth}
        \centering
        \includegraphics[width=\textwidth]{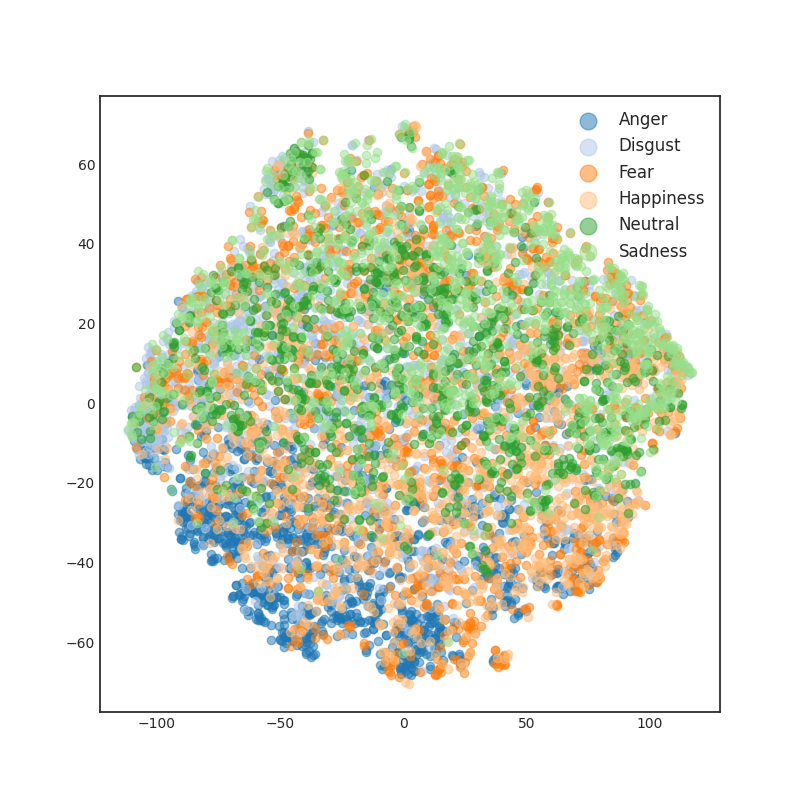}
        \caption{MMS - SER}
    \end{subfigure}
    \hfill
    \begin{subfigure}[b]{0.31\textwidth}
        \centering
        \includegraphics[width=\textwidth]{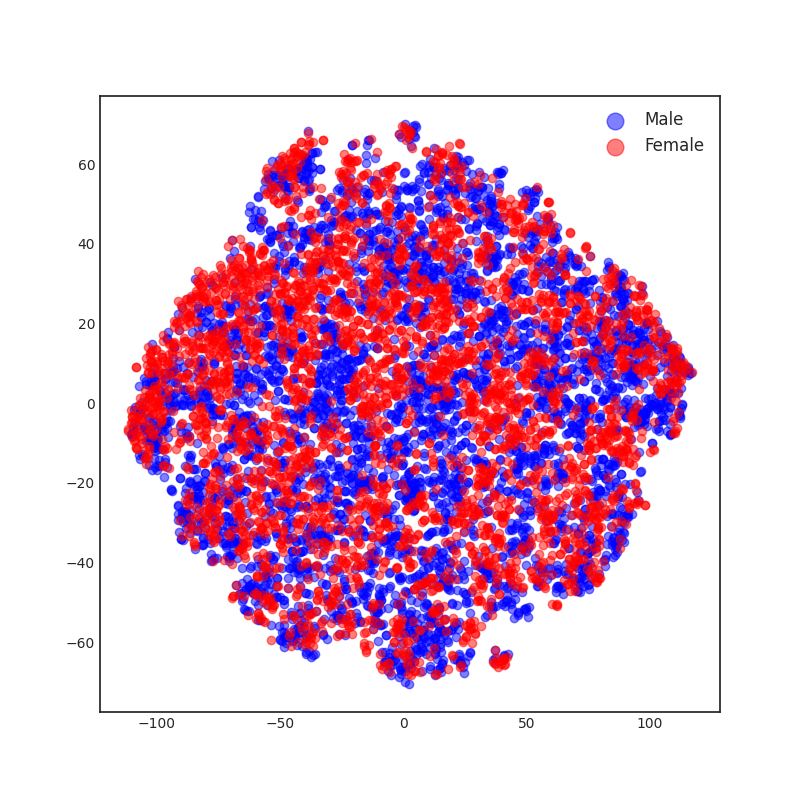}
        \caption{MMS - GR}
    \end{subfigure}
    
    \caption{t-SNE plots.}
    \label{fig:tsne1}
\end{figure*}

\begin{figure*}[htbp]
    \centering
    \begin{subfigure}[b]{0.31\textwidth}
        \centering
        \includegraphics[width=\textwidth]{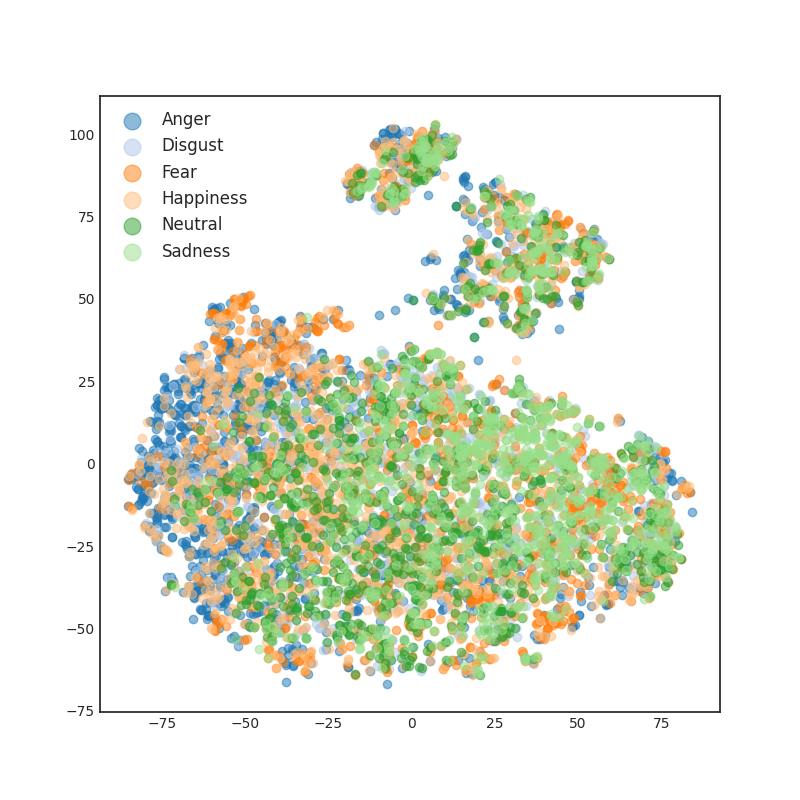}
        \caption{MFCC - SER}
    \end{subfigure}
    \hfill
    \begin{subfigure}[b]{0.31\textwidth}
        \centering
        \includegraphics[width=\textwidth]{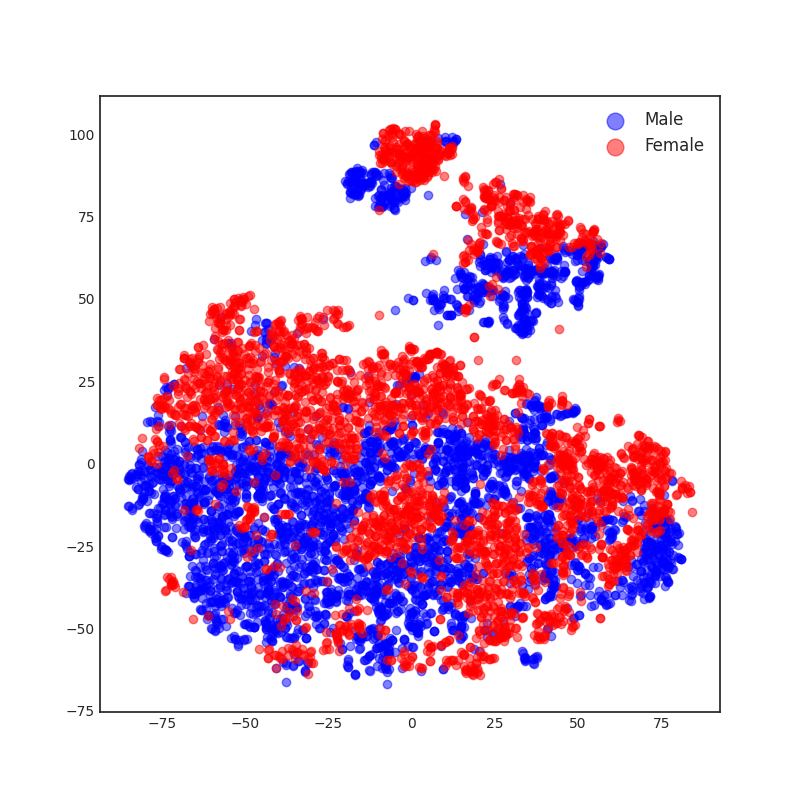}
        \caption{MFCC - GR}
    \end{subfigure}
    \hfill
    \begin{subfigure}[b]{0.31\textwidth}
        \centering
        \includegraphics[width=\textwidth]{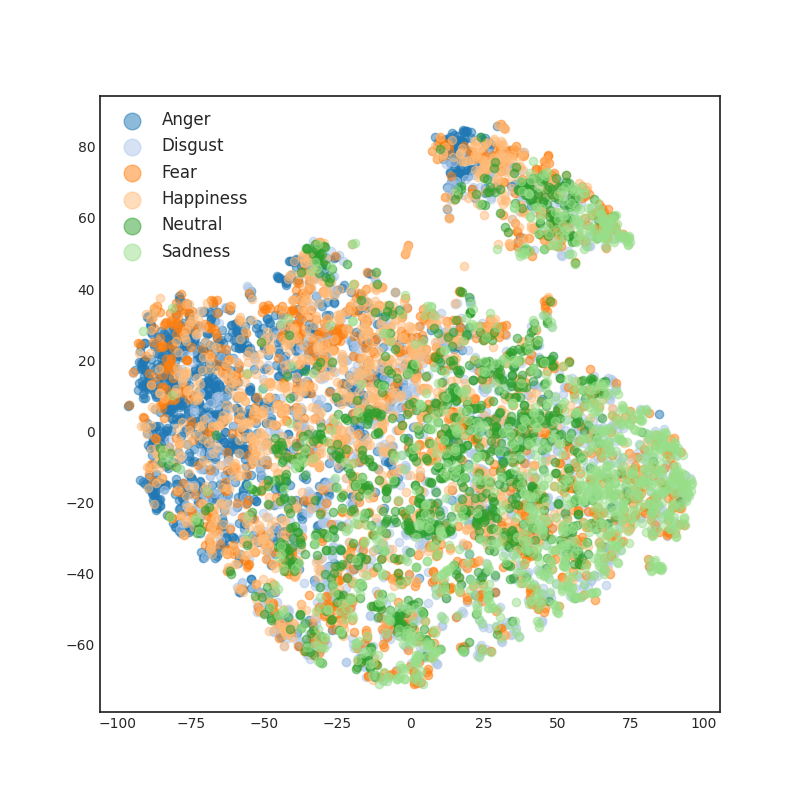}
        \caption{Wav2Vec2 - SER}
    \end{subfigure}
    
    \vspace{0.5cm} 

    \begin{subfigure}[b]{0.31\textwidth}
        \centering
        \includegraphics[width=\textwidth]{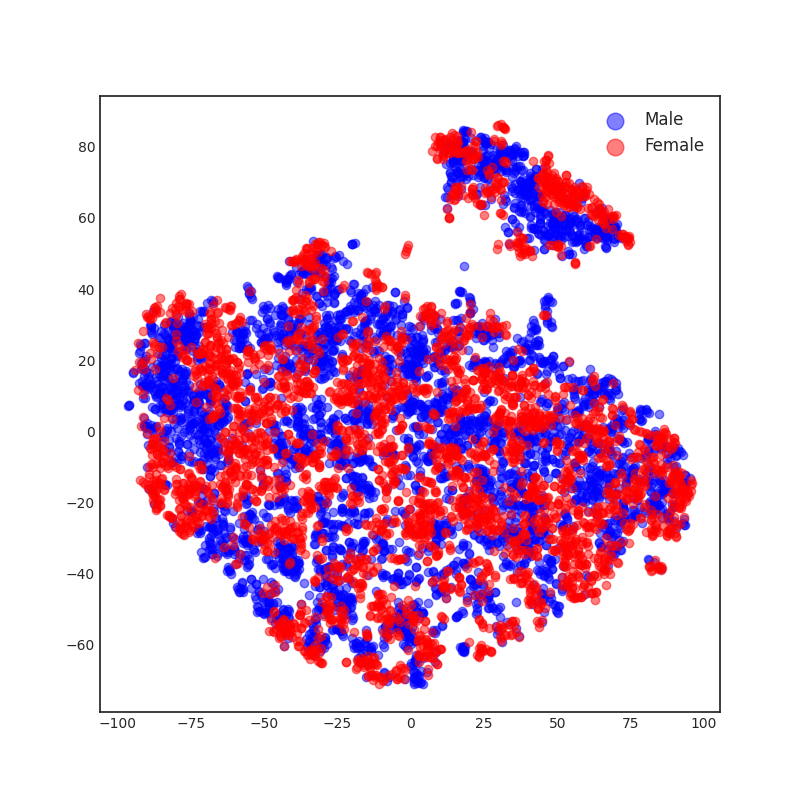}
        \caption{Wav2Vec2 - GR}
    \end{subfigure}
    \hfill
    \begin{subfigure}[b]{0.31\textwidth}
        \centering
        \includegraphics[width=\textwidth]{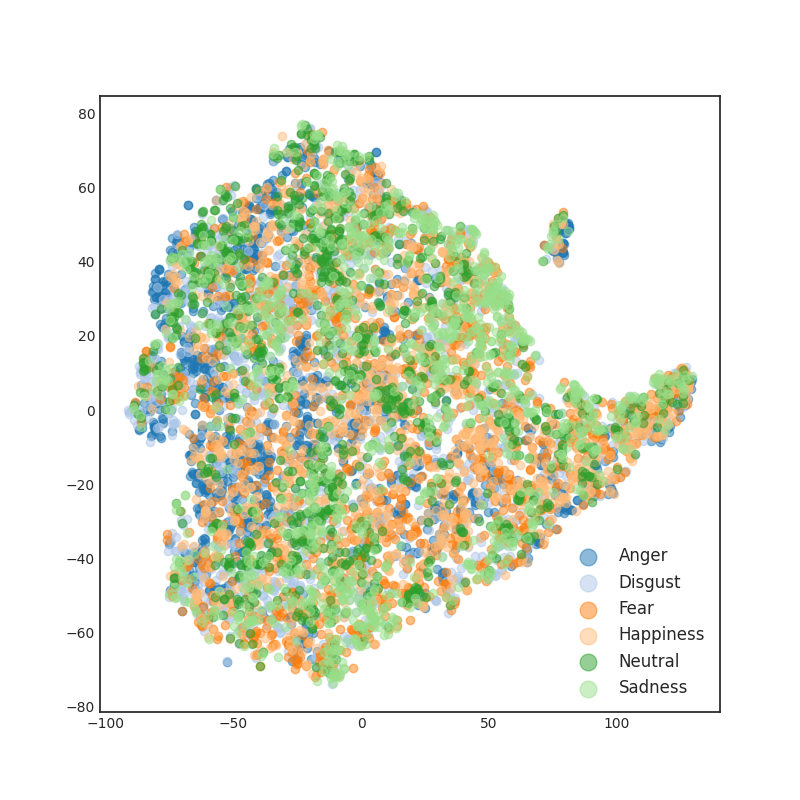}
        \caption{XLS-R - SER}
    \end{subfigure}
    \hfill
    \begin{subfigure}[b]{0.31\textwidth}
        \centering
        \includegraphics[width=\textwidth]{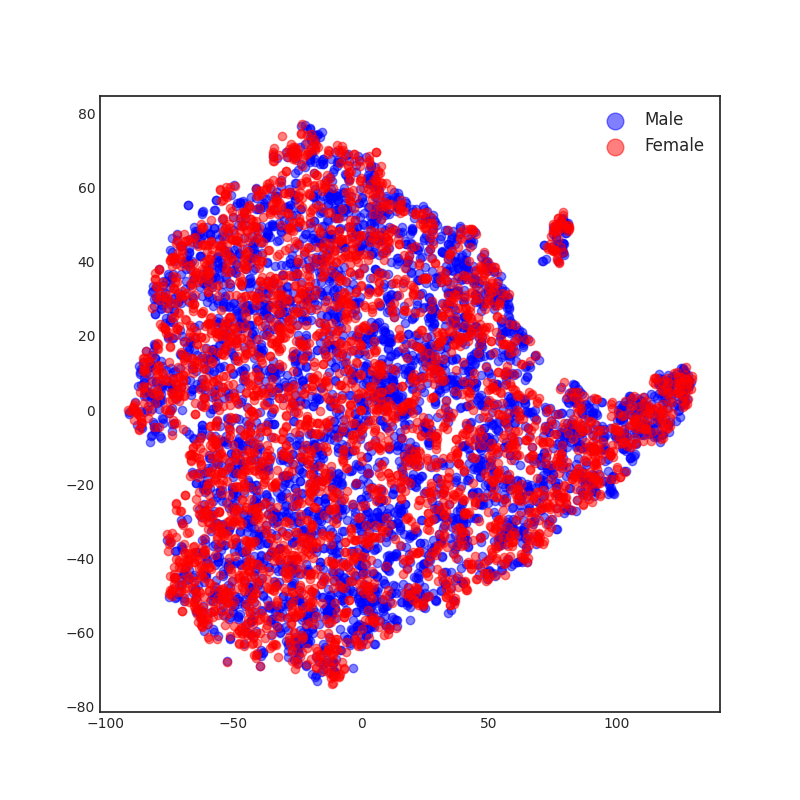}
        \caption{XLS-R - GR}
    \end{subfigure}
    
    \caption{t-SNE plots.}
    \label{fig:tsne2}
\end{figure*}

We consider accuracy as an evaluation metric for ASR, SER, and GR; RMSE for AE following previous research \cite{zheng22b_interspeech, shor2022universal}. All the models are evaluated in a 5-fold manner, and we present the average scores of 5-fold. The greater the accuracy values and the lower the RMSE, the better the models are. Table \ref{table:final_comp} summarizes the results across different models and we present the only best scores for each setup for each SFT. The evaluation results and discussion of the various models are presented below, organized by specific tasks:

\noindent \textbf{SVST}: Table \ref{table:svst} presents the results of individual SFMs for individual tasks. Models trained on TRILLsson representations outperform all the SFMs representaions by achieving the topmost performance in all the four SFTs ASR, SER, GR, and AE. This top performance of TRILLsson can be attributed to its pre-training for paralinguistic speech processing applications as the SFTs primarily depend on the paralinguistic attributes of speech such as pitch, intensity, tone, etc. 
The t-SNE plots (Appendix Figure \ref{fig:tsne1} and Figure \ref{fig:tsne2}) of raw SFM representations for SER and GR highlight the superior performance of TRILLsson, as clearer clusters are observed across gender and emotion classes.

\noindent \textbf{SVMT}: Table \ref{table:svmt} presents the results of individual SFMs representations for multi-task learning of the SFTs. Here we can observe that the performance dropped than the models trained for individual SFT. TRILLsson is still holding the topmost position, however, we observe drop than SVST. This degradation in performance can be attributed to task interference. This interference arises because each SFT depends on distinct paralinguistic features in the input speech, and individual SFMs may struggle to disentangle task-specific information effectively. 

\noindent \textbf{MVMT}: Two modeling techniques are used for MVMT, first, fusion of the views through concatenation, which we consider as a baseline fusion technique. Secondly, fusion through, \textbf{TANGO}. Table \ref{table:mvmtc}, \ref{table:mvmtb}, \ref{table:mvmte} presents the scores for MVMT results on CREMA-D, BAVED, and emo-DB for the SFTs ASR, SER, GR, and AE. Across the three datasets, we observe that fusion through \textbf{TANGO} gives better than concatenation-based baseline fusion in most instances and as well as than SVMT scenarios. These results show that MVL through \textbf{TANGO} from SFMs can mitigate task interference during multi-task learning of SFTs. Different SFM capture different aspects of the input speech signal due to different SFM's unique characteristics inherent to it, helping to disentangle task-specific information more effectively when combined, thus showing complementary behavior. This approach reduces interference by allowing each task to access the most relevant features from the appropriate view, leading to improved performance and better generalization across tasks. The combination of TRILLsson and x-vector with \textbf{TANGO} has shown the topmost performance in comparison to all the different combinations. As seen with the SVST results (Table \ref{table:svst}), better individual task SFMs, when combined, leverage their unique advantages and offer richer, more task-specific information, leading to superior overall performance. \par

We also observe an interesting phenomenon: the synergy between MFCC, despite being statistical handcrafted features, and SFMs like x-vector or TRILLsson outperforms combinations involving only MMS, UniSpeech-SAT, or other SFMs. This integration of low-level spectral features with complementary SFMs reduces redundancy and enhances noise resilience, leading to more effective representations. Overall, from these results, we can observe the effectiveness of \textbf{TANGO} for aligning views from different SFMs, thus inducing complementary strengths of the SFMs. Additionally, we also conducted an ablation study of \textbf{TANGO}. In \textbf{TANGO}, we transport features from both the SFMs to each other and for ablation experiments, we only transport features from each SFM to other and vice versa. The evaluation results are shown in Appendix Table \ref{table:tangocc}, \ref{table:tangobc}, \ref{table:tangoec} for CREMA-D, BAVED, emo-DB respectively, however, we observe one direction transportation is not able to outerperform \textbf{TANGO}.

\vspace{-0.2cm}

\section{Conclusion}\label{chapter:Conclusion and Future Work}

In this work, for the first time, we gave an comprehensive comparative investigation of various SOTA SFMs for learning SFT such as ASR, SER, GR, and  AE simultaneously as training individuals models for each task comes with resource, time, cost, and maintenence challenges. We conducted our experiments on three benchmark datasets consisting of CREMA-D, BAVED, and emo-DB. We show that multitask learning of SFTs with individual SFMs representations is prone to task interference thus degration in performance across the tasks and as a remedy, we propose, MVL. MVL aims to exploit the complementary behavior of different SFMs unique abstract space for effective learning of the SFTs parallely. For better MVL, we propose, \textbf{TANGO}, a multi-task learning framework that leverages OT as fusion mechanism. Through our experiments, we show that combination of TRILLsson and x-vector SFMs coupled with \textbf{TANGO} attains far better performance than the individual SFMs representations, and as well as the baseline fusion techniques for learning SFTs concurrenty. 

\section{Limitations}

CREMA-D, BAVED, and emo-DB were the only easily openly available datasets that contain these four tasks' information, so we only experimented with them. We experimented with only CNN as the downstream network, and this may limit our study, as previous research has shown that the downstream behavior of the SFMs changes according to the downstream network chosen \cite{zaiem23b_interspeech}. So, we plan to extend our study, by experimenting with different downstream networks. \par

One more limitation is the selection of optimal values for the task-specific loss in the total multi-task learning loss function. In this study, we have tried with only one combination, however, different values can give different results maybe better. so, in future, we plan to come up with learnable loss function, that learns the weights of the losses dynamically.

\section{Ethics Statement}
We use openly accessible datasets, and no sensitive user information was mentioned in the datasets.
\bibliographystyle{acl_natbib}
\bibliography{custom}

\section{Appendix}

\begin{table}[hbt!]
\scriptsize
  \centering
  \begin{tabular}{l|c|c|c|c}
    \toprule
    \textbf{SFMs} & \textbf{ASR} & \textbf{SER} & \textbf{GR} & \textbf{AE} \\
    \midrule
    \multicolumn{5}{c}{\textbf{\textit{TANGO(x1->x2)}}} \\
    \midrule
    XLS-R + MMS & 29.68 & 33.04 & 96.98 & 10.07 \\
    XLS-R + Whisper & 34.18 & 43.18 & 94.56 &  10.67\\
    XLS-R + Unispeech-SAT & 59.91 & 58.23 & 97.45 & 9.34 \\
    XLS-R + MFCC & 69.78 &  44.33 & 89.52 & 7.07 \\
    XLS-R + Wav2vec2 & 59.03 & 47.55 & 96.98 & 9.46 \\
    XLS-R + x-vector & 75.69 & 53.12 & 98.93 & 7.45 \\
    XLS-R + TRILLsson & 81.93 & 74.82 & 99.13 & 6.54 \\
    XLS-R + WavLM & 60.64 &  62.32 & 97.92 & 8.85 \\
    MMS + Whisper & 30.96 & 35.33 & 93.49 & 10.70 \\
    MMS + Unispeech-SAT & 32.30 & 37.27 & 95.97 & 9.86 \\
    MMS + MFCC & 52.79 & 30.49 & 87.37 & 9.02 \\
    MMS + Wav2Vec2 & 35.53 & 38.75 & 89.52 & 7.63 \\
    MMS + x-vector & 70.45 & 43.25 & 98.19 & 10.32 \\
    MMS + TRILLsson & 64.41 & 69.11 & 97.45 & 7.17 \\
    MMS + WavLM & 45.53 & 60.85 & 98.32 & 9.78 \\
    Whisper + Unispeech-SAT & 50.03 & 57.42 & 98.05 & 9.65 \\
    Whisper + MFCC & 64.81 & 39.42 & 98.93 & 8.24 \\
    Whisper + Wav2Vec2 & 41.97 & 45.53 & 96.17 & 10.92 \\
    Whisper + x-vector & 74.21 & 56.68 & 97.85 & 7.70 \\
    Whisper + TRILLsson & 81.40 & 72.33 & 99.33 & 7.25 \\
    Whisper + WavLM & 50.50 & 61.45  & 97.04 & 9.23\\
    Unispeech-SAT + MFCC & 82.34 & 62.19 & 97.78 & 6.78 \\
    Unispeech-SAT + Wav2Vec2 & 63.94 & 62.86 & 97.85 & 8.62 \\
    Unispeech-SAT + x-vector & 81.53 & 61.92 & 98.39 & 6.83 \\
    Unispeech-SAT + TRILLsson & 83.01 & 73.47 & 99.19  & 6.36 \\
    Unispeech-SAT + WavLM & 63.06 & 65.35 & 97.78 & 8.82 \\
    MFCC + Wav2Vec2 & 83.41 & 48.69 & 98.25 & 7.94 \\
    MFCC + x-vector & 71.46 & 41.37 & 96.91 & 10.01 \\
    MFCC + TRILLsson & 88.31 & 75.02 & 99.26 & 5.75 \\
    MFCC + WavLM & 84.15 & 60.17 & 98.25 & 6.93 \\
    Wav2Vec2 + x-vector & 77.77 & 55.14 & 98.39 & 7.26 \\
    Wav2Vec2 + TRILLsson & 82.94 & 73.00 & 99.33 & 6.49  \\
    Wav2Vec2 + WavLM & 67.29 & 63.47 & 98.12 & 8.80 \\
    x-vector + TRILLsson & 86.23 & 74.75 & 98.32 & 6.09 \\
    x-vector + WavLM & 82.14  & 62.19 & 98.66 & 7.40 \\
    TRILLsson + WavLM&  84.22 & 73.63 & 99.26 & 6.40 \\
    \midrule
    \multicolumn{5}{c}{\textbf{\textit{TANGO(x2->x1)}}} \\
    \midrule
    XLS-R + MMS & 33.65 & 34.65 & 94.22 & 9.75 \\
    XLS-R + Whisper & 42.24 & 43.05 & 94.90 & 10.70 \\
    XLS-R + Unispeech-SAT & 58.43 & 61.18 & 98.05 & 9.56 \\
    XLS-R + MFCC & 79.99 & 50.71 & 96.78 & 7.22 \\
    XLS-R + Wav2vec2 & 60.31 & 51.31 & 97.72 & 9.13 \\
    XLS-R + x-vector & 76.29 & 55.88 & 98.59 &  7.29 \\
    XLS-R + TRILLsson & 83.08 & 74.01 & 98.66 & 6.70 \\
    XLS-R + WavLM & 60.44 & 64.47 & 97.31 & 8.77 \\
    MMS + Whisper & 28.95 & 43.18 & 94.90 & 9.81  \\
    MMS + Unispeech-SAT & 40.56 & 61.11 &98.05  & 9.39\\
    MMS + MFCC & 12.09 & 29.15 & 82.14 & 13.26 \\
    MMS + Wav2Vec2 & 56.01 & 56.21 & 97.78 &  9.16 \\
    MMS + x-vector & 56.95 & 34.05 & 96.98 & 7.82 \\
    MMS + TRILLsson &  78.04 & 73.20 & 99.06 & 7.31 \\
    MMS + WavLM & 56.21 & 54.00 & 98.66 & 8.33 \\
    Whisper + Unispeech-SAT & 48.69 & 59.37 & 96.04 & 10.52 \\
    Whisper + MFCC &  59.97 & 46.94 & 96.98 & 8.65 \\
    Whisper + Wav2Vec2 & 43.25 & 57.29 & 96.31 & 9.41 \\
    Whisper + x-vector & 77.23 & 56.35 & 98.66 & 7.05 \\
    Whisper + TRILLsson & 79.18 & 74.35 & 99.26 & 7.30 \\
    Whisper + WavLM & 51.04 & 61.79 & 97.78 & 9.78 \\
    Unispeech-SAT + MFCC & 84.35 & 64.14 & 98.39 & 7.14 \\
    Unispeech-SAT + Wav2Vec2 & 64.00 & 63.00 & 97.85  & 8.64 \\
    Unispeech-SAT + x-vector & 82.87 & 65.21 & 98.39 & 6.82 \\
    Unispeech-SAT + TRILLsson & 83.68 & 74.75 & 99.33 & 6.55 \\
    Unispeech-SAT + WavLM & 64.61 & 66.82 & 97.72 & 8.76 \\
    MFCC + Wav2Vec2 & 84.69 & 48.89 & 97.99 & 6.90 \\
    MFCC + x-vector & 88.11 & 58.23 & 99.13 & 6.05 \\
    MFCC + TRILLsson & 90.13 & 75.35 & 98.86 & 5.71 \\
    MFCC + WavLM & 84.75 & 61.38 & 98.25 & 6.94 \\
    Wav2Vec2 + x-vector & 81.20 & 55.14 & 98.32 & 7.29 \\
    Wav2Vec2 + TRILLsson & 81.93 & 74.68 & 99.46 & 6.62 \\
    Wav2Vec2 + WavLM & 67.23 & 64.88 &97.99  &  8.78\\
    x-vector + TRILLsson & 87.04 & 74.35 & 99.19 & 6.00 \\
    x-vector + WavLM & 82.07 & 63.33 & 98.59 & 6.94 \\
    TRILLsson + WavLM& 83.61 & 73.47 & 99.06 & 6.53 \\
    \bottomrule
  \end{tabular}
  \caption{Accuracy (ASR, SER, GR) in percentage and RMSE (AE) values for different SFMs in Multi-View Multi-Task Setup for CREMA-D.}
  \label{table:tangocc}
\end{table}

\begin{table}[hbt!]
\scriptsize
  \centering
  \begin{tabular}{l|c|c|c|c}
    \toprule
    \textbf{SFMs} & \textbf{ASR} & \textbf{SER} & \textbf{GR} & \textbf{AE} \\
    \midrule
    \multicolumn{5}{c}{\textbf{\textit{TANGO(x1->x2)}}} \\
    \midrule
    XLS-R + MMS & 86.30  & 78.55  & 98.45 & 3.60 \\
    XLS-R + Whisper & 76.49  & 81.65  & 96.12 & 2.86 \\
    XLS-R + Unispeech-SAT & 85.53 & 79.07 & 98.19 & 3.60 \\
    XLS-R + MFCC & 83.98 & 80.62 & 98.97 & 3.42 \\
    XLS-R + Wav2vec2 & 84.75 & 81.65 & 98.71 & 4.01 \\
    XLS-R + x-vector & 87.86 & 83.20 & 98.97 & 2.56 \\
    XLS-R + TRILLsson & 89.66 & 86.82 & 99.74 & 1.74 \\
    XLS-R + WavLM & 85.27 & 78.04 & 9948 & 2.76 \\
    MMS + Whisper & 73.90 & 82.95 & 99.22 & 3.45 \\
    MMS + Unispeech-SAT & 81.65 & 78.81 & 97.67 & 3.85 \\
    MMS + MFCC & 85.27 & 78.81 & 98.97 & 4.05 \\
    MMS + Wav2Vec2 & 80.88 & 76.74 & 98.71 & 3.18 \\
    MMS + x-vector & 86.30 & 83.72 & 99.74 & 2.74 \\
    MMS + TRILLsson & 88.37 & 84.75 & 99.74 & 1.90 \\
    MMS + WavLM & 80.36 & 84.24 & 94.83 & 2.41 \\
    Whisper + Unispeech-SAT & 84.24 & 80.36 & 99.48 & 2.44 \\
    Whisper + MFCC & 89.66 & 78.04 & 99.74 & 2.80 \\
    Whisper + Wav2Vec2 & 82.43 & 78.55 & 99.48 & 3.40 \\
    Whisper + x-vector & 83.20 & 78.04 & 100.00 & 3.32 \\
    Whisper + TRILLsson & 90.70 & 85.27 & 100.00 & 1.71 \\
    Whisper + WavLM & 84.50 & 78.29 & 99.74 & 3.24 \\
    Unispeech-SAT + MFCC & 90.44 & 82.17 & 98.19 & 2.77 \\
    Unispeech-SAT + Wav2Vec2 & 85.01 & 77.26 & 98.71 & 3.38 \\
    Unispeech-SAT + x-vector & 87.86 & 82.17 & 99.74 & 2.12 \\
    Unispeech-SAT + TRILLsson & 88.37 & 86.05 & 99.74 & 2.65 \\
    Unispeech-SAT + WavLM & 86.56 & 82.17 & 99.74 & 1.87 \\
    MFCC + Wav2Vec2 & 88.37 & 80.88 & 98.97 & 2.62 \\
    MFCC + x-vector & 90.25 & 83.98 & 99.74 & 2.50 \\
    MFCC + TRILLsson & 89.92 & 87.86 & 99.74 & 2.01 \\
    MFCC + WavLM & 89.41 & 79.84 & 99.22 & 2.01 \\
    Wav2Vec2 + x-vector & 88.63 & 83.46 & 99.74 & 2.13 \\
    Wav2Vec2 + TRILLsson & 89.41 & 86.05 & 99.74 & 2.35 \\
    Wav2Vec2 + WavLM & 86.56 & 83.20 & 99.22 & 2.95 \\
    x-vector + TRILLsson & 89.66 & 85.15 & 100.00 & 1.90 \\
    x-vector + WavLM & 86.56 & 81.40 & 100.00 & 1.95 \\
    TRILLsson + WavLM & 90.44 & 84.24 & 100.00 & 2.07 \\
    \midrule
    \multicolumn{5}{c}{\textbf{\textit{TANGO(x2->x1)}}} \\
    \midrule
     XLS-R + MMS & 82.95 & 83.72 & 97.42 & 3.56 \\
     XLS-R + Whisper & 72.09 & 69.25 & 97.67 & 4.99 \\
     XLS-R + Unispeech-SAT & 88.37 & 78.04 & 98.97 & 3.22 \\
     XLS-R + MFCC & 86.05 & 83.46 & 97.16 & 3.54 \\
     XLS-R + Wav2vec2 & 87.60 & 81.14 & 97.93 & 4.41 \\
     XLS-R + x-vector & 87.34 & 82.69 & 99.48 & 2.28 \\
     XLS-R + TRILLsson & 88.11 & 84.41 & 99.48 & 2.11 \\
     XLS-R + WavLM & 85.53 & 81.14 & 99.48 & 3.19 \\
     MMS + Whisper & 83.20 & 85.27 & 99.22 & 4.32 \\
     MMS + Unispeech-SAT & 87.08 & 76.49 & 99.74 & 1.89 \\
     MMS + MFCC & 80.62 & 74.68 & 98.97 & 4.82 \\
     MMS + Wav2Vec2 & 81.91 & 81.91 & 98.45 & 2.89 \\
     MMS + x-vector & 86.56 & 81.65 & 98.71 & 2.66 \\
     MMS + TRILLsson & 89.66 & 84.75 & 100.00 & 1.98 \\
     MMS + WavLM & 86.05 & 81.14 & 98.19 & 2.66 \\
     Whisper + Unispeech-SAT & 78.04 & 78.29 & 99.48 & 3.00 \\
     Whisper + MFCC & 84.24 & 64.86 & 98.19 & 4.68 \\
     Whisper + Wav2Vec2 & 85.53 & 78.81 & 99.48 & 2.92 \\
     Whisper + x-vector & 86.30 & 84.50 & 99.48 & 2.51 \\
     Whisper + TRILLsson & 87.08 & 85.27 & 100.00 & 1.78 \\
     Whisper + WavLM & 86.56 & 81.40 & 98.97 & 3.20 \\
     Unispeech-SAT + MFCC & 88.89 & 78.81 & 99.74 & 2.89 \\
     Unispeech-SAT + Wav2Vec2 & 85.01 & 79.07 & 98.97 & 3.52 \\
     Unispeech-SAT + x-vector & 83.20 & 81.91 & 99.74 & 2.39 \\
     Unispeech-SAT + TRILLsson & 90.44 & 84.75 & 100.00 & 1.86 \\
     Unispeech-SAT + WavLM & 87.60 & 79.59 & 100.00 & 2.65 \\
     MFCC + Wav2Vec2 & 89.66 & 82.17 & 98.71 & 2.50 \\
     MFCC + x-vector & 90.47 & 82.69 & 100.00 & 1.87 \\
     MFCC + TRILLsson & 90.44 & 86.05 & 99.22 & 2.26 \\
     MFCC + WavLM & 90.51 & 83.98 & 99.74 & 2.46 \\
     Wav2Vec2 + x-vector & 90.96 & 84.24 & 99.74 & 2.20 \\
     Wav2Vec2 + TRILLsson & 90.96 & 86.30 & 100.00 & 1.91 \\
     Wav2Vec2 + WavLM & 89.66 & 79.84 & 99.74 & 2.63 \\
     x-vector + TRILLsson & 88.89 & 85.27 & 99.22 & 2.14 \\
     x-vector + WavLM & 89.92 & 81.65 & 100.00 & 2.22 \\
     TRILLsson + WavLM & 90.18 & 83.34 & 100.00 & 1.79 \\
    \bottomrule
  \end{tabular}
  \caption{Accuracy (ASR, SER, GR) in percentage and RMSE (AE) values for different SFMs in Multi-View Multi-Task Setup for BAVED.}
  \label{table:tangobc}
\end{table}

\begin{table}[hbt!]
\scriptsize
  \centering
  \begin{tabular}{l|c|c|c|c}
    \toprule
    \textbf{SFMs} & \textbf{ASR} & \textbf{SER} & \textbf{GR} & \textbf{AE} \\
    \midrule
    \multicolumn{5}{c}{\textbf{\textit{TANGO(x1->x2)}}} \\
    \midrule
    XLS-R + MMS & 71.03 & 85.05 & 95.33 & 2.58 \\
    XLS-R + Whisper & 82.24 & 82.24 & 96.26 & 2.85 \\
    XLS-R + Unispeech-SAT & 83.18 & 93.46 & 98.13 & 2.46 \\
    XLS-R + MFCC & 91.59 & 85.05 & 97.20 & 2.93 \\
    XLS-R + Wav2vec2 & 88.79 & 89.72 & 97.20 & 2.05 \\
    XLS-R + x-vector & 91.59 & 88.79 & 99.07 & 2.06 \\
    XLS-R + TRILLsson & 96.26 & 96.26 & 100.00 & 1.75 \\
    XLS-R + WavLM & 91.59 & 92.52 & 100.00 & 1.84 \\
    MMS + Whisper & 74.77 & 76.64 & 100.00 & 3.12 \\
    MMS + Unispeech-SAT & 78.50 & 90.65 & 97.20 & 2.33 \\
    MMS + MFCC & 82.24 & 82.24 & 97.20 & 2.67 \\
    MMS + Wav2Vec2 & 76.64 & 76.64 & 97.20 & 2.52 \\
    MMS + x-vector & 94.39 & 85.98 & 99.07 & 1.86 \\
    MMS + TRILLsson & 97.20 & 97.20 & 100.00 & 1.73 \\
    MMS + WavLM & 92.52 & 94.39 & 100.00 & 2.64 \\
    Whisper + Unispeech-SAT & 83.18 & 89.72 & 98.13 & 2.47 \\
    Whisper + MFCC & 85.98 & 83.18 & 99.07 & 2.57 \\
    Whisper + Wav2Vec2 & 74.77 & 85.05 & 98.13 & 2.98 \\
    Whisper + x-vector & 96.26 & 94.39 & 100.00 & 1.42 \\
    Whisper + TRILLsson & 99.07 & 98.13 & 100.00 & 1.63 \\
    Whisper + WavLM & 90.65 & 89.72 & 100.00 & 2.00 \\
    Unispeech-SAT + MFCC & 92.52 & 85.05 & 98.13 & 2.06 \\
    Unispeech-SAT + Wav2Vec2 & 92.52 & 91.59 & 98.13 & 2.44 \\
    Unispeech-SAT + x-vector & 95.33 & 91.59 & 98.13 & 1.95 \\
    Unispeech-SAT + TRILLsson & 95.33 & 96.26 & 100.00 & 1.78 \\
    Unispeech-SAT + WavLM & 87.85 & 92.52 & 99.07 & 2.15 \\
    MFCC + Wav2Vec2 & 92.52 & 91.59 & 99.07 & 2.57 \\
    MFCC + x-vector & 97.20 & 90.65 & 100.00 & 1.81 \\
    MFCC + TRILLsson & 97.20 & 97.20 & 99.07 & 1.66 \\
    MFCC + WavLM & 96.26 & 89.72 & 99.07 & 2.72 \\
    Wav2Vec2 + x-vector & 96.26 & 92.52 & 100.00 & 1.56 \\
    Wav2Vec2 + TRILLsson & 98.13 & 97.20 & 100.00 & 1.78 \\
    Wav2Vec2 + WavLM & 95.33 & 91.59 & 100.00 & 2.07 \\
    x-vector + TRILLsson & 97.20 & 99.07 & 100.00 & 1.74 \\
    x-vector + WavLM & 96.26 & 93.46 & 100.00 & 1.72 \\
    TRILLsson + WavLM & 98.23 & 100.00 & 100.00 & 1.48 \\
    \midrule
    \multicolumn{5}{c}{\textbf{\textit{TANGO(x2->x1)}}} \\
    \midrule
    XLS-R + MMS & 73.83 & 86.92 & 96.26 & 2.60 \\
    XLS-R + Whisper & 71.96 & 83.18 & 100.00 & 2.78 \\
    XLS-R + Unispeech-SAT & 89.72 & 90.65 & 99.07 & 2.65 \\
    XLS-R + MFCC & 85.05 & 84.11 & 97.20 & 2.83 \\
    XLS-R + Wav2vec2 & 83.18 & 92.52 & 98.13 & 2.81 \\
    XLS-R + x-vector & 94.39 & 92.52 & 100.00 & 1.76 \\
    XLS-R + TRILLsson & 98.13 & 100.00 & 100.00 & 1.64 \\
    XLS-R + WavLM & 87.85 & 89.72 & 100.00 & 2.15 \\
    MMS + Whisper & 71.96 & 85.05 & 95.33 & 3.13 \\
    MMS + Unispeech-SAT & 77.57 & 95.33 & 97.20 & 2.46 \\
    MMS + MFCC & 87.85 & 83.18 & 97.20 & 2.78 \\
    MMS + Wav2Vec2 & 89.72 & 74.77 & 100.00 & 2.67 \\
    MMS + x-vector & 98.13 & 92.52 & 100.00 & 1.75 \\
    MMS + TRILLsson & 99.07 & 97.20 & 100.00 & 1.64 \\
    MMS + WavLM & 96.26 & 92.52 & 99.07 & 1.93 \\
    Whisper + Unispeech-SAT & 81.31 & 82.24 & 99.07 & 2.54 \\
    Whisper + MFCC & 85.98 & 88.79 & 99.07 & 2.13 \\
    Whisper + Wav2Vec2 & 78.50 & 87.85 & 100.00 & 2.71 \\
    Whisper + x-vector & 98.13 & 94.39 & 100.00 & 2.18 \\
    Whisper + TRILLsson & 97.20 & 98.13 & 100.00 & 1.93 \\
    Whisper + WavLM & 91.59 & 87.85 & 99.07 & 2.41 \\
    Unispeech-SAT + MFCC & 93.46 & 90.65 & 99.07 & 2.10 \\
    Unispeech-SAT + Wav2Vec2 & 95.33 & 91.59 & 99.07 & 2.46 \\
    Unispeech-SAT + x-vector & 97.20 & 96.26 & 100.00 & 1.45 \\
    Unispeech-SAT + TRILLsson & 99.07 & 99.07 & 100.00 & 1.49 \\
    Unispeech-SAT + WavLM & 90.65 & 94.39 & 100.00 & 2.10 \\
    MFCC + Wav2Vec2 & 95.33 & 86.92 & 99.07 & 2.42 \\
    MFCC + x-vector & 97.20 & 91.59 & 100.00 & 1.43 \\
    MFCC + TRILLsson & 98.13 & 96.26 & 100.00 & 1.58 \\
    MFCC + WavLM & 92.52 & 90.65 & 99.07 & 2.08 \\
    Wav2Vec2 + x-vector & 96.26 & 91.59 & 100.00 & 2.09 \\
    Wav2Vec2 + TRILLsson & 99.07 & 100.00 & 100.00 & 1.81 \\
    Wav2Vec2 + WavLM & 93.46 & 88.79 & 100.00 & 1.88 \\
    x-vector + TRILLsson & 98.13 & 94.39 & 100.00 & 1.69 \\
    x-vector + WavLM & 96.26 & 92.52 & 100.00 & 1.85 \\
    TRILLsson + WavLM & 98.02 & 98.13 & 100.00 & 1.54 \\
    \bottomrule
  \end{tabular}
  \caption{Accuracy (ASR, SER, GR) in percentage and RMSE (AE) values for different SFMs in Multi-View Multi-Task Setup for emo-DB.}
  \label{table:tangoec}
\end{table}

\subsection{Hyperparameter and Training details}
\label{sec:hyper}

We keep the convolutional block 1D-CNN kernel to be of size 3 with 32 and 64 as number of kernels. We keep the number of neurons as 200, 64, 56 for SVST setup. For SVMT, MVMT, \textbf{TANGO}, for the task-specific head we keep the number of neurons as 30. we use softmax in the output for the models for ASR, SER and GR tasks. We use linear activation function for AE. We use Adam as the optimizer and set the learning rate to 1e-3 with a batch size of 32.

\begin{figure}[bt] 
        \centering
        \includegraphics[width=0.475\textwidth]{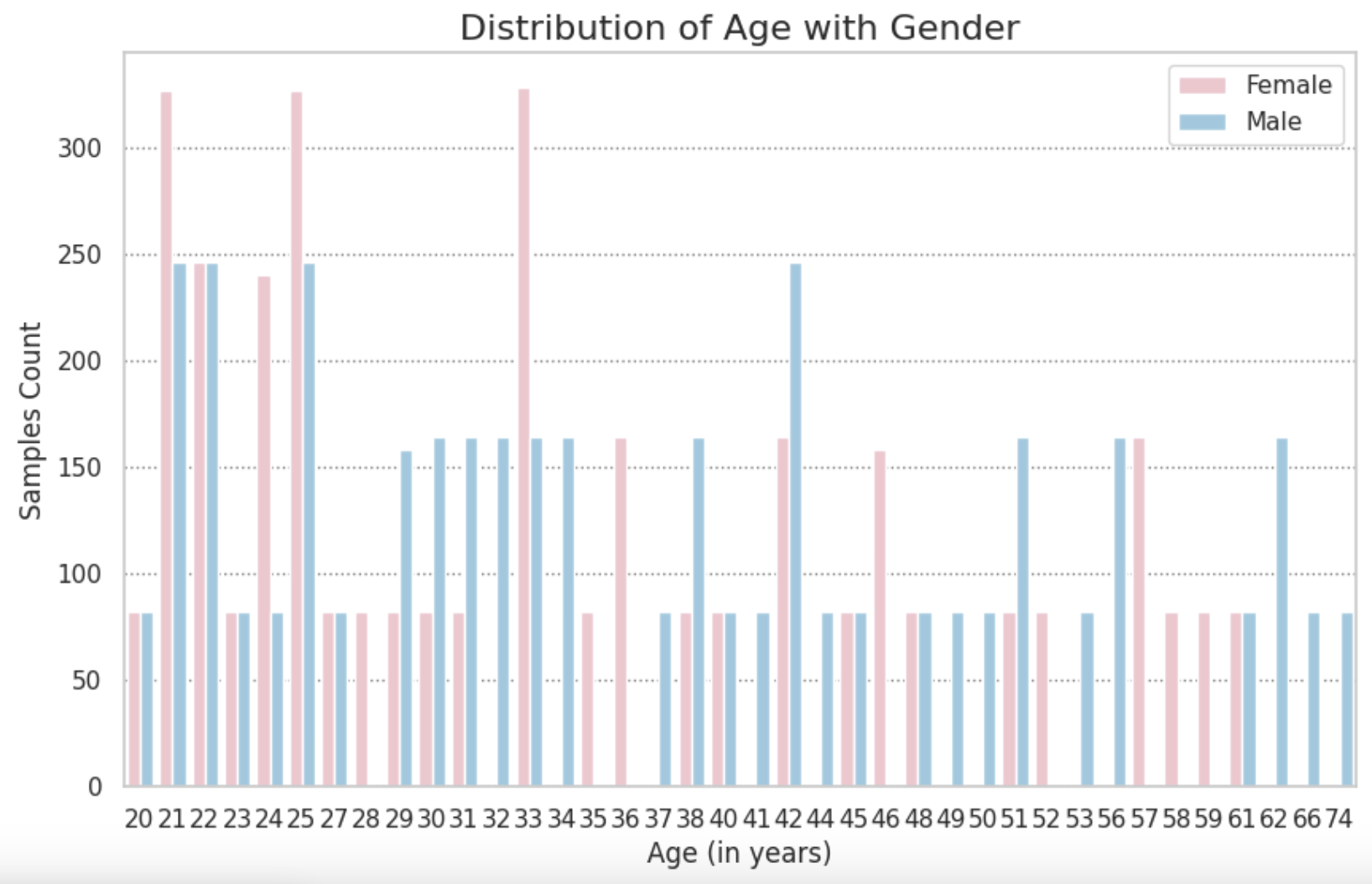} 
        \caption{Age and Gender Distribution for CREMA-D.}
        \label{fig:cremadageemo}
\end{figure}

\begin{figure}[bt] 
        \centering
        \includegraphics[width=0.475\textwidth]{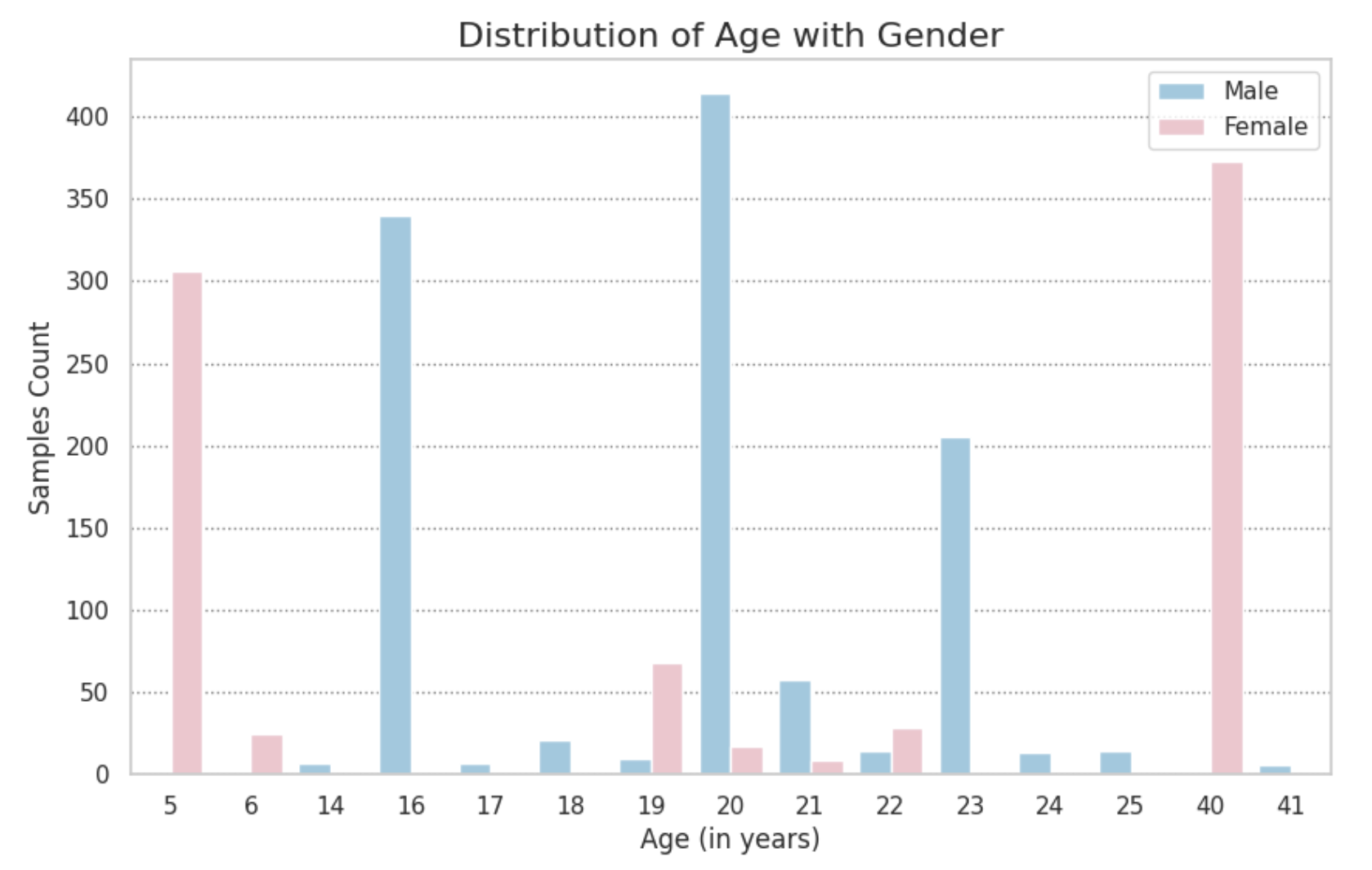} 
        \caption{Age and Gender Distribution for BAVED.}
        \label{fig:bavedageemo}
\end{figure}

\end{document}